# Resource Management and Quality of Service Provisioning in 5G Cellular Networks

Sam Kayyali*

***Abstract*— With the commercial launch of 5G technologies and fast pace of expansion of cellular network infrastructure, it is expected that cellular and mobile networks traffic will exponentially increase. In addition, new services are expected to spread widely, such as the Internet of Things connected to mobile networks. This will add additional burden in terms of traffic load. As a result, some studies suggest that mobile traffic may increase more than 1000 times compared to the amount of traffic that is generated nowadays. This means that network resources for mobile services must be managed and controlled in a smart way, because resources are always limited, but the demand for services and the need for keeping user equipment always connected to mobile networks can be considered unlimited, leaving gap between huge service demands and available resources. In order to narrow this gap, major consideration should be given to the management of network resources to avoid network congestion and performance degradation during peak hour/s and traffic spikes, and allow access to network services to more customers when demand is high. On the other hand, guaranteeing quality of service requirements for the wide range of new services is another challenge that must be met in 5G networks. In this paper we will review 5G networks characteristics and specifications, then carry out a survey on resource management and QoS provisioning to improve and manage resource utilization in 5G networks.**

## I. INTRODUCTION

It is obvious that the number of mobile subscribers and users has been boosting over the last decade. In 2018, the number of mobile devices exceeded two billion [1]. The growth of mobile networks devices and subscribers means that the generated traffic will also rise at unprecedented rates forcing mobile network operators (MNO) to adopt new solutions, techniques, and technologies to fill the gap between the increasing demand on cellular networks data services and the available capacity of existing networks [2]. In Fig. 1, we can see the huge increase in monthly traffic of smart mobile devices and connections, and the growth of mobile devices in the period between 2017 – 2022 [3].

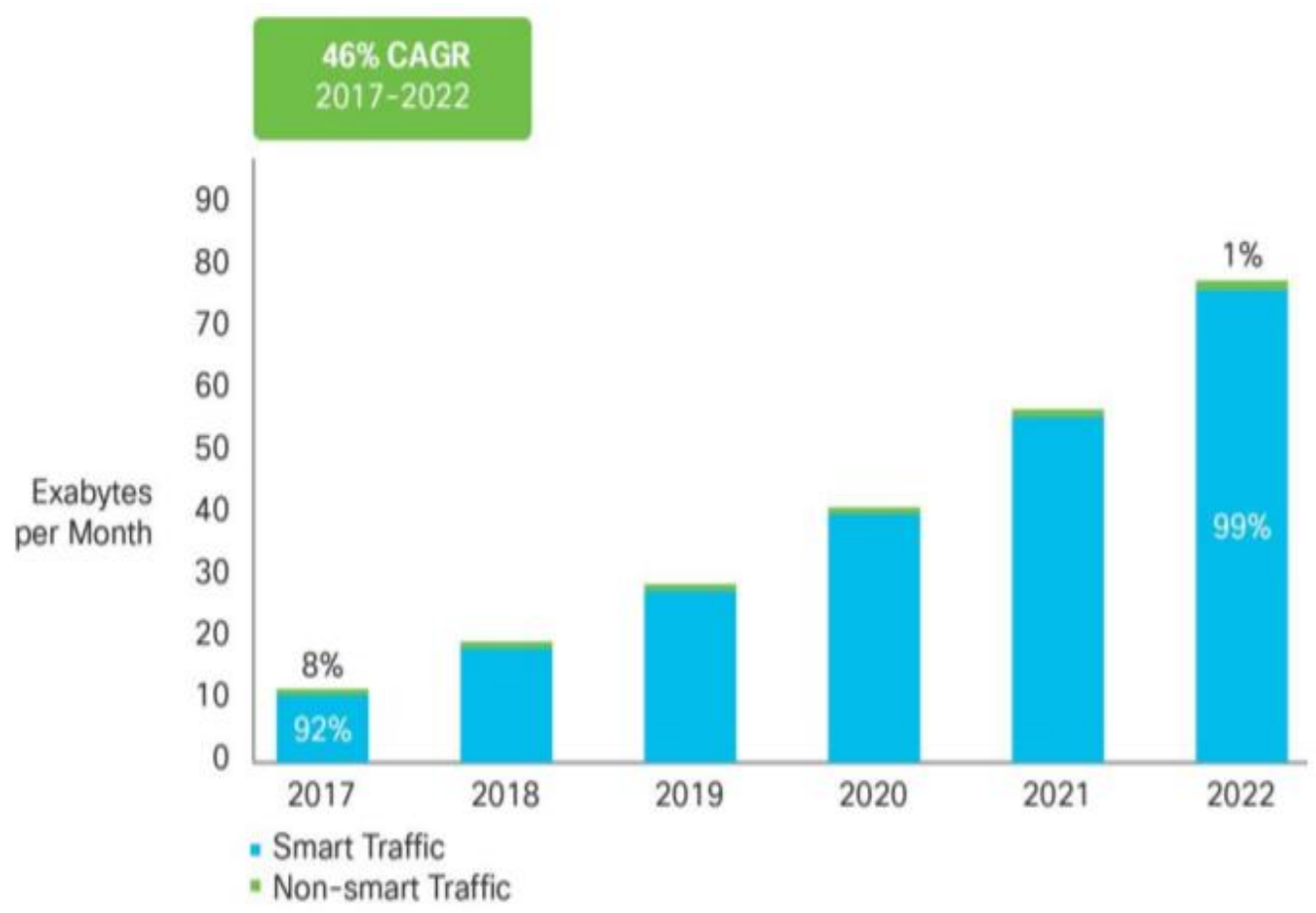

Fig. 1. (a) Mobile Traffic Growth (cisco VNI mobile, 2019)

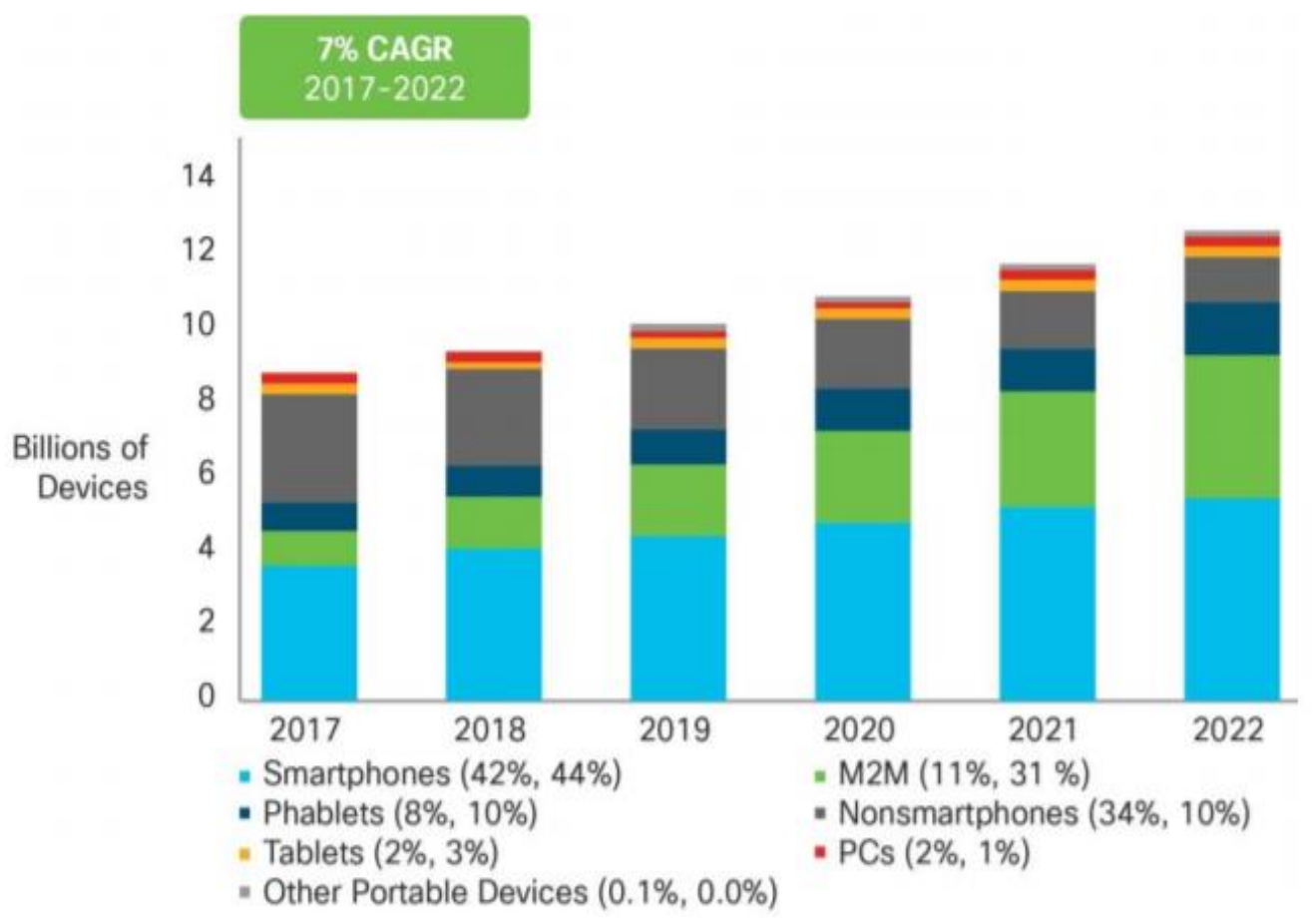

Fig. 1. (b) Mobile devices Growth (cisco VNI mobile, 2019)

In the same context, the spread of Covid-19 has added additional burden on mobile and wireless networks as a result of social distancing restrictions. During this pandemic, almost all social, educational, economical, and government activities have become more dependable on technological advances including wireless and cellular technologies than physical methods. In addition, the fact that the contribution of mobile phone data in investigating the spread of covid-19 virus could have been more effective with the adoption of advanced technologies (5G mobile generation) on wide range over the world during this pandemic [4], [5] should also be taken into our consideration. We need to look at these factors as lessons learned and motivations to shift toward the new 5G networks that promise efficient technologies to fulfil current and future requirements in telecommunication industry. To guarantee a smooth transition to the 5th generation of mobile networks, it is very important to tackle the core features of 5G mobile networks and research their impact on the performance to avoid unwanted key performance indicator (KPI) degradation that will certainly impact user experience [6]. One major area of study of 5G networks is resource management schemes and their contribution to improving the overall performance of cellular network through the allocation of 5G network resources [7] based on different criteria and techniques that we are going to study. The rest of the paper includes four main sections. In section II, we will cover the existing commercial cellular networks, their major features, and their limitations. In section III, we will go through the main characteristics and advances of 5G networks over older technologies like LTE. Section IV will cover main network techniques of 5G networks. In section V of the paper will focus on resource management schemes in 5G wireless cellular networks. Section VI will discuss the contributions of quality of service (QoS) provisioning schemes in 5G networks. Finally, section VII will be the conclusion and future work.

* School of Information Technology, Carleton University
Ottawa - Canada
e-mail: sam.kayyali.ca@gmail.com

## II. Background on cellular systems (2G, 3G, 4G)

With the commercial launch of mobile services in 1991 [8], the world has been in feverish race to develop methods to enhance cellular applications, specifically data services, as the need for data is increasing along with sufficient data rates that have been evolving over the years to meet business and users requirements.

Second generation technology (2G – 2.75G), offered data rates up to 384 Kbps with the introduction of Enhanced data rates for GSM evolution (EDGE) [9]. The main disadvantages of this system were quality, capacity, global roaming, and the limitation of the amount of data that can be sent [10].

Third generation technology of mobile (3G - 3.75G) was able to achieve up to 14.4 Mbps on the downlink with High Speed Downlink Packet Access (HSDPA), and a throughput on the uplink up to 5.76 Mbps by using High Speed Uplink Packet Switched (HSUPA) [11].

Fourth generation became wireless standard in 2009 and achieved a peak throughput of more than 100 Mbps [12], and 50 Mbps in the uplink [11] [13]. Advanced long term evolution (LTE-A) was able to better fulfil smartphone users requirements compared to second and third generations and new applications and services appeared as a result of the advantages of higher data rates and better spectrum efficiency that LTE and LTE-A brought to cellular users [13]. Main contributor to the performance leap of fourth generation mobile networks was the use of orthogonal frequency division multiple access (OFDMA) that divides the bandwidth of the carrier frequency into sub-carriers and each user is allocated a subset of the sub-carriers [14]. In general, third generation (3G) and fourth generation (4G) systems focus on improving metrics like spectrum efficiency, peak data throughput, and coverage [15], but they have major challenges that still need to be addressed including optimum allocation of available spectrum and resources which has improved with 4G, but yet, it is still a limitation in 4G cellular net-works. Another challenge is the huge consumption of energy in 4G and 3G networks, leading to more emissions of carbon-di-oxide. In addition, 4G technology deployment almost reached to its maximum capacity [16], and there is no way to improving spectral efficiency significantly despite of all trials to optimize system performance and radio links [17].

With these limitations and challenges in existing cellular networks, we can say that these systems will not be able to achieve the target of increasing data rates to more than 10 Gbps with low latency, nor to cater the forecasted increase of data traffic, and reducing green gases emissions for healthy environment [16]. In addition, there are other factors limiting existing systems like the inefficient search for relevant information in the huge amount of available data, this mechanism must be enhanced. The concept of machine-to-machine (M2M) communications, which is currently limited to specific applications, cannot be implemented in large scale due its high requirements that existing cellular system cannot afford [18]. In next section, we will introduce literature review on fifth generation mobile networks and how the features of 5G technology can help to overcome the limitations and challenges in older systems.

## III. LITERATURE REVIEW ON 5G CELLULAR NETWORKS

As discussed in section (I), the limitations in existing cellular networks and the high expectations of mobile and wireless services are the motivations for shifting to 5G technology that will achieve much higher data rates, lower delays, better coverage, lower energy consumption, and improved quality of service. To meet this vision, developments to some known technologies used in existing cellular systems should be considered and new strategy for designing 5G networks architecture should also be adopted. The literature review will cover two parts, the first part (A) demonstrates the requirements that must be met by 5G networks, then in the second part (B), we will review some new technologies and developments in 5G.

### *A. 5G Networks Specifications:*

There are many requirements that 5G system needs to provide to realize future technology aspirations. TABLE I

shows 5G networks specifications.

TABLE. I
Specifications of 5G Networks

| Specification | Value | Description | Ref. |
|---|---|---|---|
| Latency | 0.25 ms | For specific applications like machine tools operation | [19] |
| | 1 ms | Robotics & Telepresence / VR / Health Care / Gaming | |
| | 5 ms | Education & Culture | |
| | 10 ms | Intelligent Transport System (ITS) | |
| Reliability | 99.999 % | Higher reliability is needed for critical services like Health care, V2V | [20] |
| Data Rate | 1 – 10 Gbps | VR / Gaming / Education | [19] |
| Mobility | Up to 500 Km/h | Speed supported by 5G networks is much higher than 4G (250 Km/h) | [21] |
| Connected Devices | >10 -100 times more | Much more devices of IoT, M2M, Sensors | [22] |
| Band | 3 – 300 GHZ | Unutilized higher band for more spectrum, higher throughput, better spectral efficiency. | [23] |
| Carrier bandwidth | 5 – 10 – 20 – 40 – 100 MHZ | Higher carrier bandwidth than LTE (up to 20 MHZ) for higher data rates | [24] |
| Transmission Time Interval TTI | 0.2 – 4 ms | TTI is adjustable per scheduling instant (Not fixed as in LTE) | [21], [24] |
| Spectral Efficiency (SE), Energy Efficiency (EE) | 10 times improvement | Improved SE results in better EE. Advances in MIMO, plays important role in improving SE and EE | [25] |

### B. *Developments and New Technologies in 5G*

1) *MmWave in 5G networks*: To cater the expected 1000-fold increase in traffic in the coming decade, the band in the range between 30 – 300 GHZ, which is called Extremely High Frequency (EHF) [23], can be used in 5G networks. This will add around 200 times more spectrum than the current allocated spectrum for broadband applications upon utilizing only 40% of the proposed mmWave band [17]. Despite the huge capacity that mmWave is intended to provide, there are some real concerns related to attenuation (ex. Rain), low penetration capability, and higher sensitivity to blockages [26], [23]. On the other hand, the utilization of EHF, allows the packing of more antenna elements into limited area [27], opening the door for massive MIMO technology.
2) *Massive MIMO*: Massive MIMO benefits from the fact that it is possible to fit more antenna elements in small space due to the shorter wavelength in 5G band [28]. It also increases the elements in antenna array to hundreds that can transmit over several antennas [29] and used in both the base station and the user equipment. Massive MIMO is considered a development of traditional MIMO used in LTE but with more advances to comply with 5G technology It is more spectrum and energy efficient because of using spatial multiplexing technique resulting in higher system capacity with 10 times more concurrent users who can be served in the same time frequency slot compared to LTE system, it also improves robustness against fading, noise, and failure, and it reduces latency in the air interface [30].
3) *MmWave beamforming*: To guarantee an efficient utilization of mmWave spectrum, mmWave beamforming is developed to achieve gains in cellular network. It is used to concentrate both transmitted and received signals in desired direction to minimize pathloss impact on cellular networks [31]. This is another positive impact of having more antenna elements in limited space to facilitate highly directional beams that reduces massive MIMO power level [32].
4) *Heterogeneous Networks (HetNet):* In 5G, planning of network nodes follows anapproach that is based on deploying network base stations (BSs) that have different coverage sizes and footprints. There will be macro base stations or High-Power Nodes (HPN) for rural and remote areas, and Low Power Nodes (LPN) to extend the coverage of HPN, enhance data rates, absorb traffic from HPN. Deployment of heterogenous nodes can also improve EE and network capacity compared to using traditional macro sites. Tight coordination between HPN and LPN is required coordinate available radio resources of both HPN and LPN and mitigate interference from HPN to LPN to maximizes the benefits that heterogenous nodes can bring to 5G networks [33].
5) *Multiplexing*: In LTE, Orthogonal Frequency Division Multiplexing (OFDM) has shown many advantages of fading resistance, efficient computation, and protection against interference. But on the other hand, it has strict timing requirements and high side lobes making the efficiency of the bandwidth a weakness in LTE [34]. In 5G, modulation schemes are divided into two main groups, Orthogonal Multiple Access (OMA) and the Non-Orthogonal Multiple Access (NOMA). Fig. 2 shows different techniques under both OMA and NOMA [35]. Simulations have shown that new modulation schemes have lower peak-average power ratio (PAPR), smaller side lobes allowing better spectral efficiency [34]. While OMA is a simpler technology where one sub channel is assigned to a single user only, NOMA technology includes power domain (PD-NOMA) and code domain (CD-NOMA) and allows multiplexing messages of multiple users on the same sub channel, resulting in better SE, support of more connected user equipment, and lower latency [35]. In addition, it reduces signaling overhead and does not need complex synchronization [36].

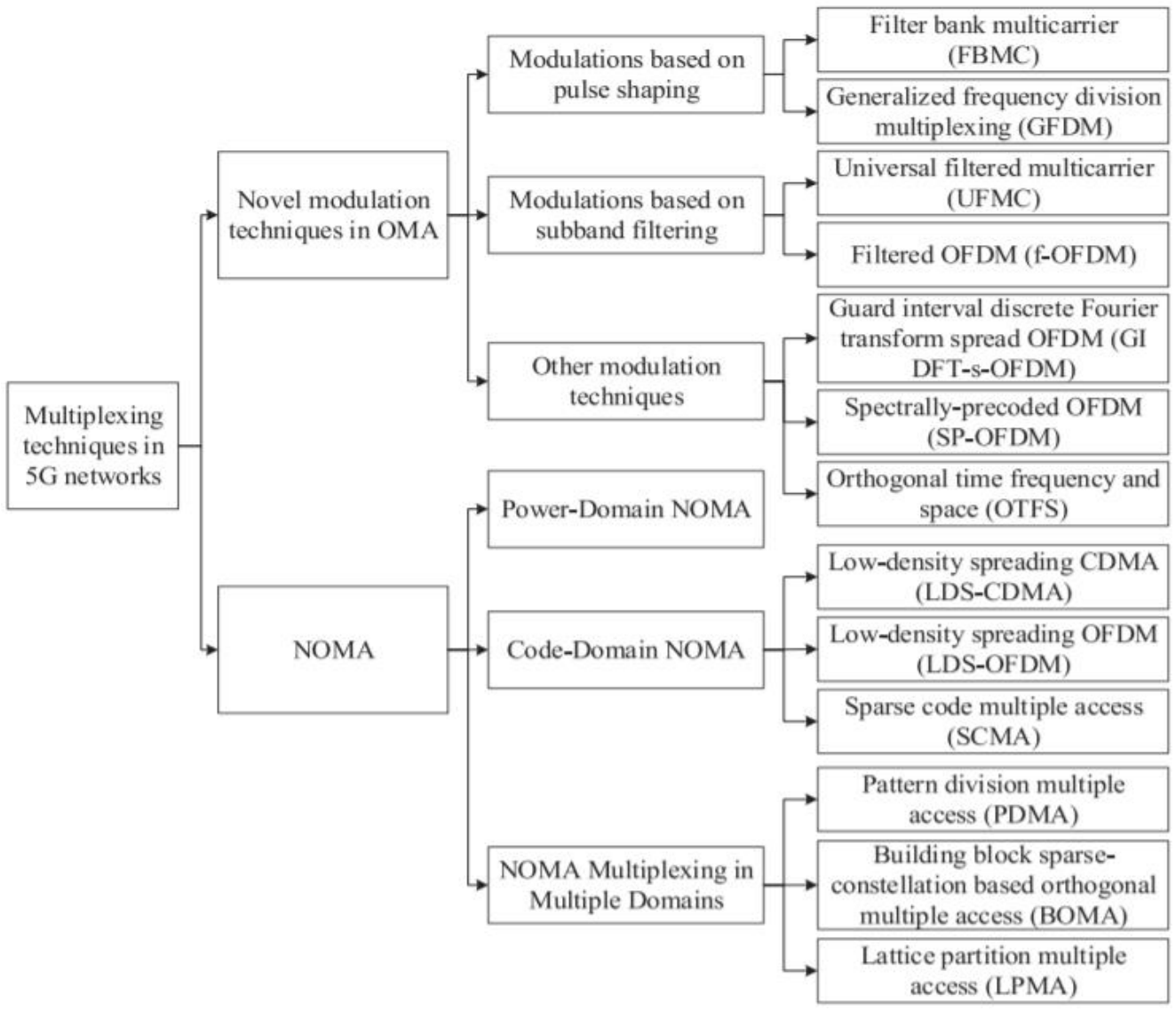

Fig. 2. Classification of 5G multiplexing techniques

6) *Device to Device communication (D2D)*: two devices can communicate with each other over the licensed cellular spectrum with limited or even no base station involvement [37]. This direct communication is more appropriate for real time services and provides lower latency and high throughput compared to the conventional communication that requires a link with a base station. Moreover, the reliability of the communication also improves. There are different applications that requires a sort of direct communication among devices such as medical devices, automotive, and IoT
7) *Network Function Virtualization (NFV) & Wireless Software Defined Networks (WSDN)*: with NFV, services can be separated from the hardware infrastructure, allowing different services to share same infrastructure for efficient resource utilization. In other words, the functionalities of network elements can be virtualized [38]. SDN network is programmable network. It separates control plane (centralized plane that holds network intelligence and controllers) from data plane that has network devices for simple packet forwarding. These devices are programable and can be controlled by central control plane for easier management for large networks. Control/user plane separation reduces the overhead of control plane, increases data rates, and improves system capacity [39]. Enhanced scalability and manageability features of NFV and SDN contribute to achieving network slicing in 5G, where physical network is sliced into separated logical networks, so that each slice has different characteristics and QoS requirements and can be dedicated accordingly to a specific service [40].
8) *Internet of Things (IoT)*: a network of physical objects, home appliances, health monitors, vehicles, buildings and more will be part of IoT. The elements of IoT network will collect and send information to a server mainly through the internet, and the server may respond back with controlling commands. New IoT applications require high performance technologies with regards to coverage of wireless communication, lower latency, and massive connectivity. Moreover, it is expected that num-bers of IoT M2M devices would exceed twenty billion by the end of 2020 [41] and more than 28 billion devices by 2021 [42], and the numbers would certainly rise with the expansion of 5G networks. These high requirements can be fulfilled by 5G new advances that would be able to create massive IoT by connecting billions of smart devices [43].
9) *Cloud Radio Access Networks (C-RAN)*: in conventional RAN, each base station BS has the transceivers or radio resource head (RRH) connected to digital processor units or baseband units (BBU) for data processing before sending to core network [44]. In C-RAN, BBU is not installed at BS anymore, BBUs are pooled in a central location [45] and RRHs of each BS is connected to the pool via fronthaul that might be a special CPRI fiber or through wireless connectivity [46] (Fig. 3 shows the difference between RAN and C-RAN). In the BBU pool, BBUs can be virtualized by a server to create virtual BBUs (v-BBU). This architecture has similar concept of cloud data centers and reduces energy consumption, provides better resource utilization and higher capacity, and interference mitigation is also improved because of BBU pooling [46].

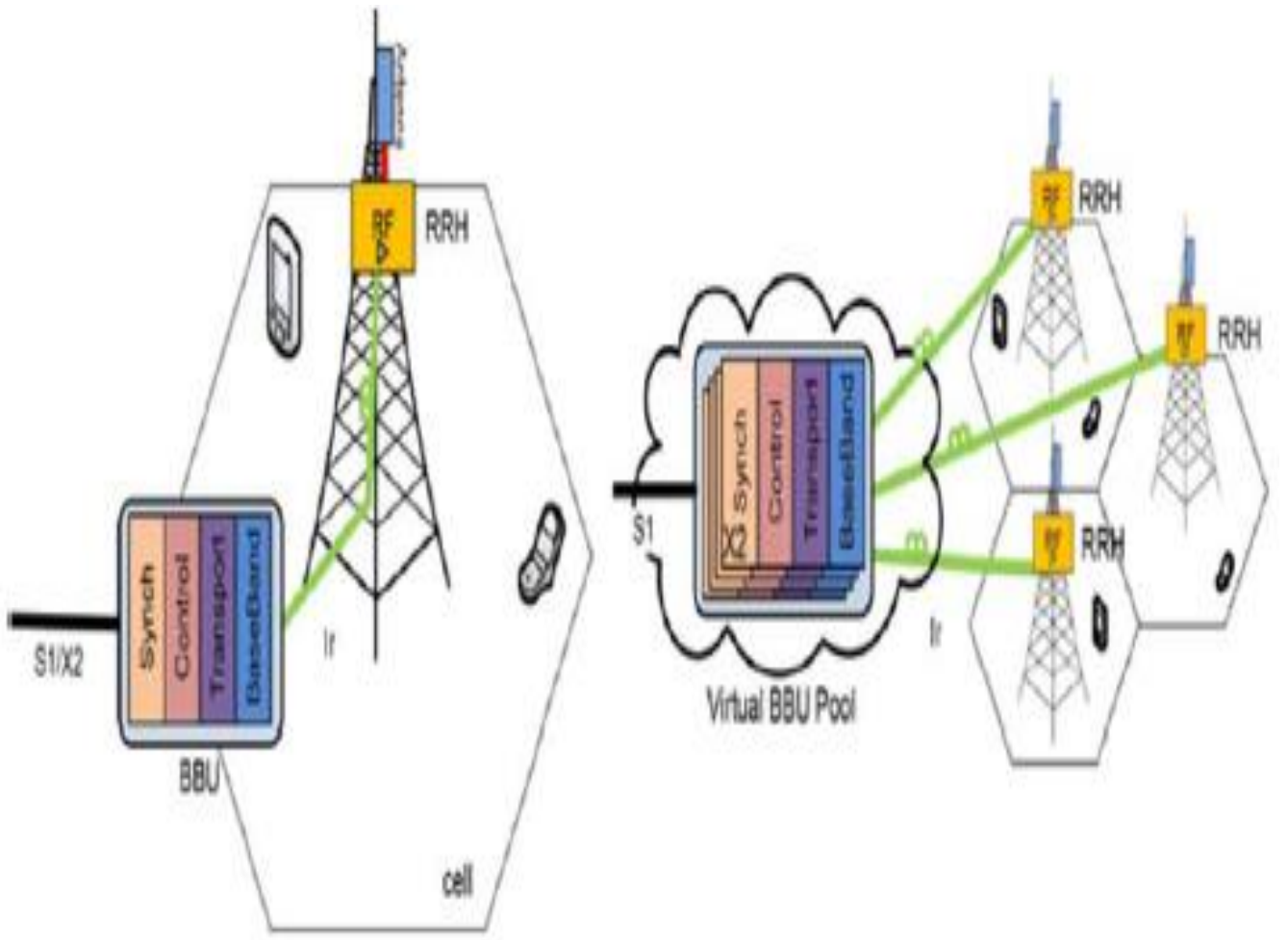

Fig. 3. Traditional RAN (left) / C-RAN (right)

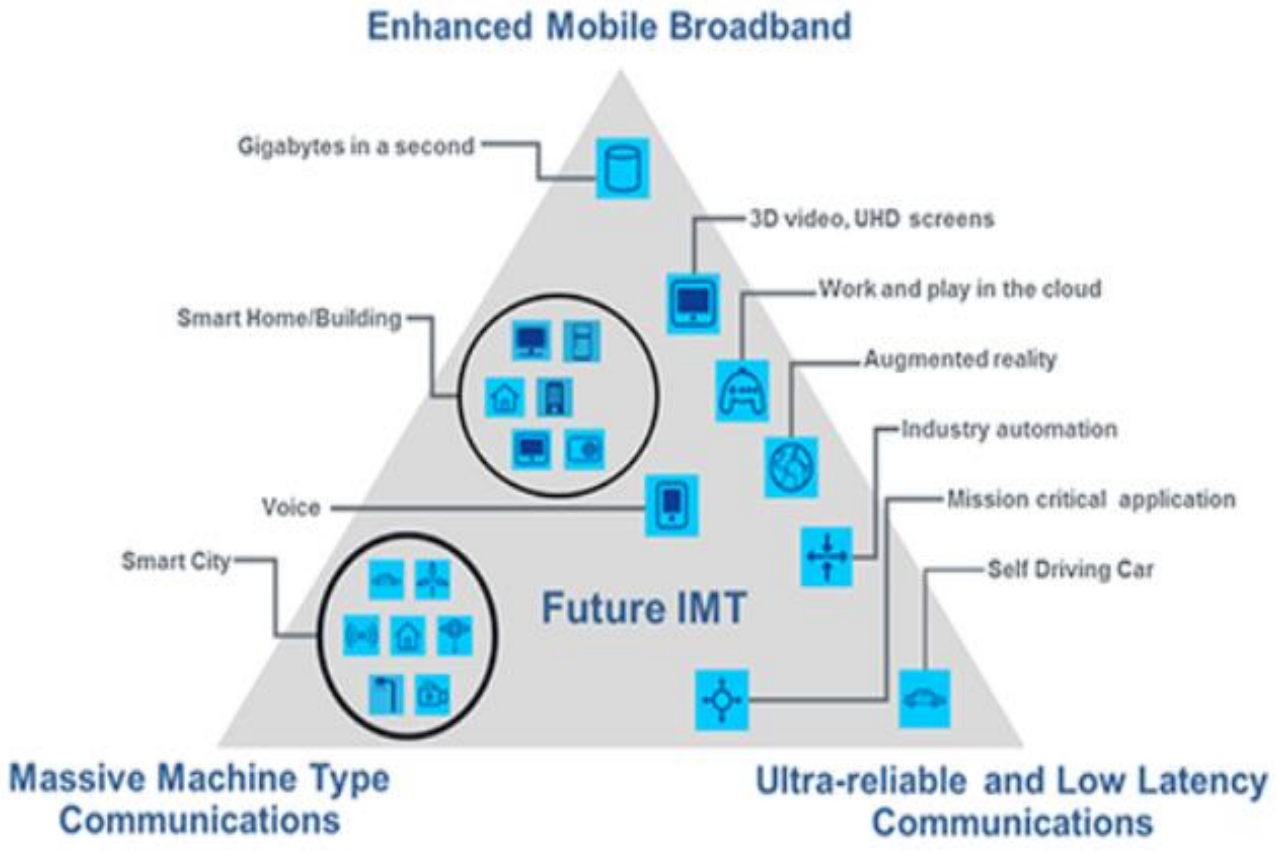

Fig. 4. (b) General and Generic Services in 5G

With all the advances brought by 5G networks, the demand on traffic will always exceed available resources. Capacity and resource management remain a major concern in mobile networks. In section IV, we will discuss resource management schemes in 5G regarding their importance and their impact on network performance, stability, and quality of service.

## IV. Networking Techniques In 5G

As discussed in section (III), 5G technology promises ambitious performance that far bypasses the performance of previous standards in terms of lower latency, higher peak data rates, and better EE, SE. We also introduced advanced and new technologies that have important role in meeting 5G requirements. But are these technologies enough to achieve the goal of 5G networks? The answer is simply NO. Since the demand for data is increasing, the available and shared resources allocated for 5G networks especially from Radio Access Network (RAN) side must be managed in a smart way so that these resources are not wasted by an inefficient resource utilization that, in turn, may cause the loss of the advantages that 5G technology is promising. As discussed earlier, 5G will pave the for new applications in multiple domains like transport, industry, health, entertainment, media, economic, automotive [48]. These applications fall in three generic service categories with different requirements: enhanced mobile broadband (eMBB) that aims to maximize throughput, massive machine-type communications (mMTC) that is characterized as huge number of communicating devices (IoT) that are able to send small amounts of data during activity periods, and ultra-reliable low-latency communications (URLLC) that requires low latency and high reliability [48] [49]. Fig. 4 shows the main services in 5G [47].

| | Latency | Mobility | Spectrum Efficiency | Data Rate | Energy Efficiency |
|---|---|---|---|---|---|
| eMBB | Med | High | High | High | High |
| uRLLC | High | High | Low | Low | Low |
| mMTC | Low | Low | Low | Low | Low |

Fig. 4. (a) Generic Services/requirements in 5G

To accommodate all these services that need heterogenous requirements, there are different methodologies that help to manage the available resources. In this survey, we will review the most important and recent papers on resource management and we will focus on RAN resource management due to the geographical distribution of RAN elements and the complexity of radio access technologies (RAT) deployed in the RAN part of the network, that require high degree of coordination to intelligently utilize the shared resources [50].

There are some technologies related to controlling, managing, allocating, and sharing resources in 5G that we need to cover before we discuss recent resource management techniques in 5G wireless networks:

A. *Network Slicing*: As discussed in previous sections, different services need different quantitative requirements regarding latency, reliability, data rate, and spectrum. In addition to the technical requirements, they need to meet other operational needs like energy and cost. Here comes the importance of having flexible and scalable network that can support the offered services [51]. While current networks adopt the idea of one physical network for all services without providing flexibility features, 5G networks benefit from the new architecture of physical layer through the softwarization and virtualization properties that can be achieved by the technologies of SDN and NFV discussed earlier in section (III). Together, SDN and NFV can provide the required flexibility, programmability, and modularity needed to create multiple virtual or logical subnetworks on top of common physical network [52]. These virtual subnetworks are called *slices*. Each logical network slice can support specific services and can be managed independently (fig. 5 shows an example of network slices on common physical layer), so that network slice should be isolated from other slices and an action taken on one slice must not negatively impact other slices, to provide guaranteed performance for each tenant (network user) even when different tenants utilize resources of virtual slices for applications and services whose requirements are conflicting [53]. Resources allocated to network slices are adjustable so that new physical resources can be added/removed, or resources can be scaled up or down through inter-slice negotiation based on the need and load of available slices, radio conditions, and geographical area to avoid negative impact on defined slices and slice tenants [54]. In addition to softwarization and virtualization properties of SDN and NFV technologies that help to customize network slices and make sure that resources are effectively utilized, network orchestration has also an important role in coordinating slice requests and managing network resources [55].

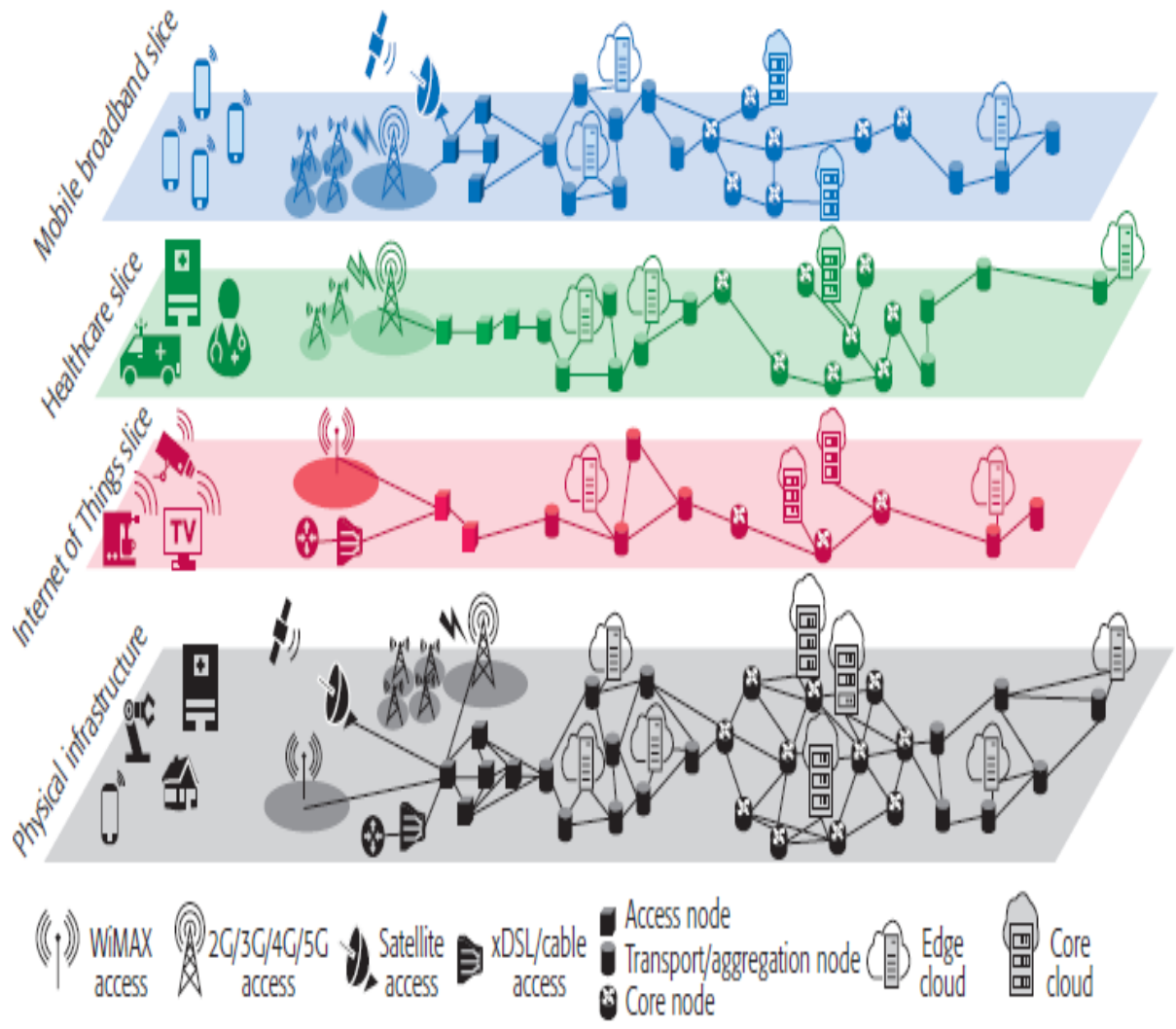

Fig. 5. 5G network slices on a common physical network

B. *Wireless Network Virtualization (WNV)*: physical resources for mobile communication including BSs (LPN, HPN), BBU, access points, Radio Access Networks (RAN), transmission links (fronthaul, backhaul), spectrum, and core network(routers, switches, gateways) owned by Mobile Network Operator (MNO) can be decoupled from the services, so that service providers (SP) can share the physical infrastructure upon abstracting and isolating physical resources to a number of virtual resources by the MNO [56], then SPs lease virtual resources and provide end-to-end services to tenants. In some cases, infrastructure provider (InP) offers the wireless network resources to be leased by mobile virtual network provider (MVNP) that creates virtual resources, then mobile virtual network operator (MVNO) operates and assigns virtual resources to SPs [57].

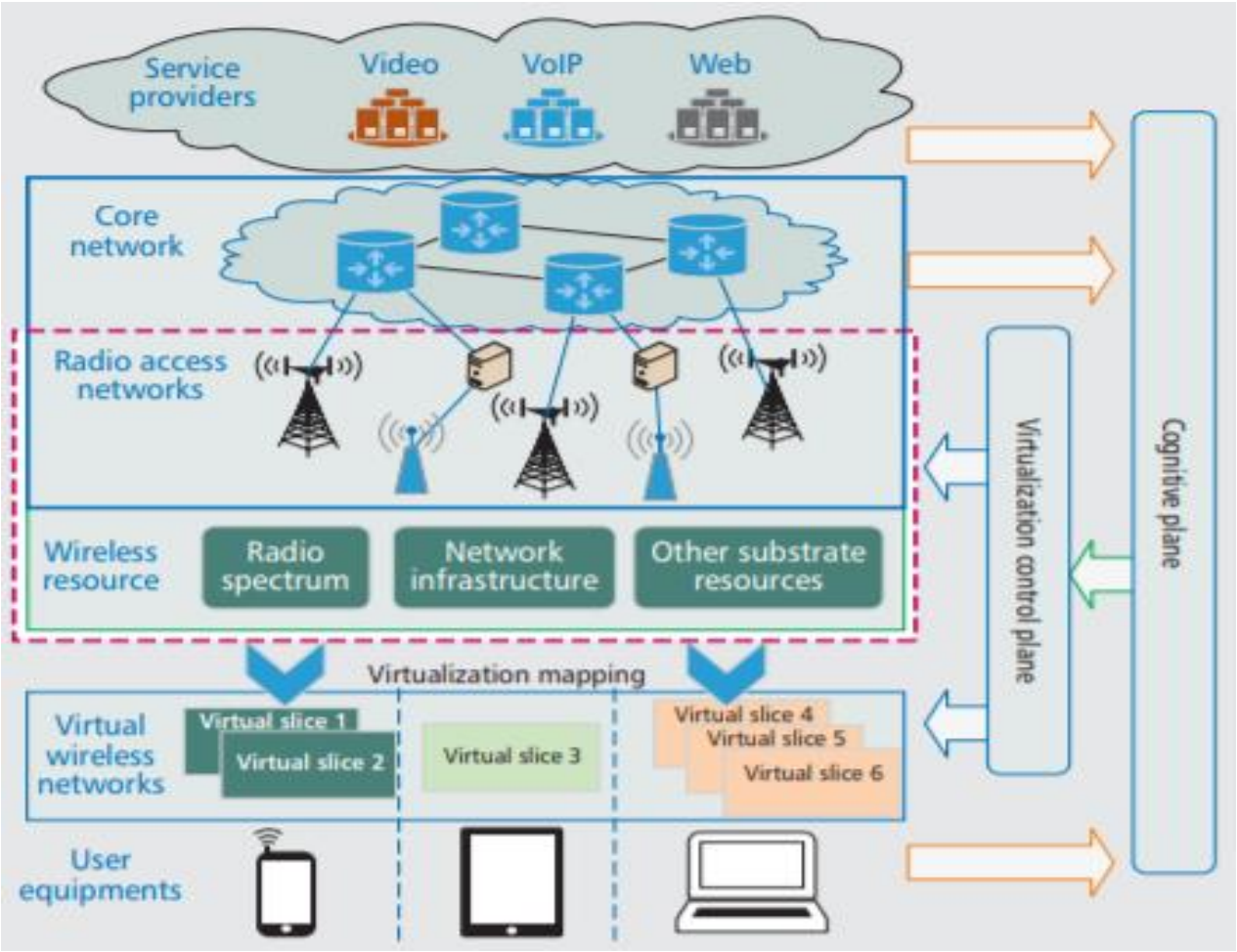

Fig. 6. Wireless network virtualization

As per [56], fig. 6 shows a model of virtualized 5G network and virtual slices of radio resources that support multiple user requirements. WNV matches network resources with user requirements by slicing the physical infrastructure into virtual resources. are virtualized for better access flexibility for network tenants and dynamic resource allocation for user equipment to meet service requirements. Traffic is transmitted by the data plane through virtual networks which are sliced from wireless resources, and the state of user requirements, services, and resources are collected by cognitive plane from SPs, UEs, and RAN to be used in network slicing and allocating resources to UEs.

C. *Network Caching*: reducing traffic in the wireless network backhaul has a huge impact on effective resource management since these backhaul transmission links (BS to core network, or between BSs) are expensive [58] [59]. It has been found that Here comes the importance of caching techniques in reducing backhaul traffic [60]. If popular contents on the internet are repeatedly requested by multiple users, in this case duplicate data will be sent over the backhaul to users over and over again, and this in turn, results in excessive load on the transmission links. Cashing means that webpages and video contents can be stored in devices installed near network edge. If BSs cache popular contents in a way that users can reach the content from BS without repeated transmission of same data over the backhaul links. Such scenario requires sending the traffic only one time via links, then BS stores the content to share it directly with multiple users, lowering backhaul transmission requirements especially during peak hours [61]. In addition to traffic reduction over the backhaul links, the cost of installing caching devices at the BS is much less than expanding backhaul transmission links to accommodate the duplicated and matching contents from multiple users [62].

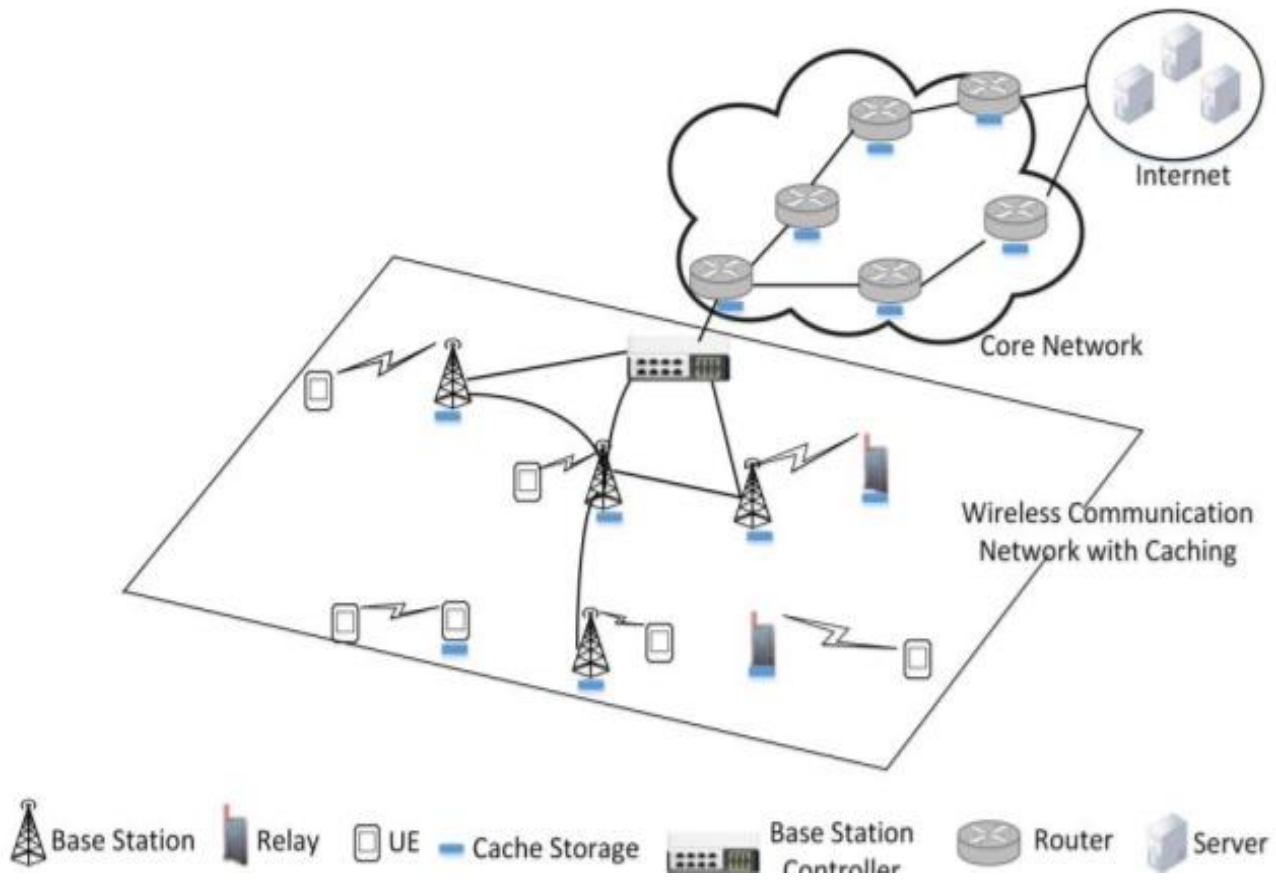

Fig. 7. Cellular network architecture with caching

D. *EDGE/Fog Computing*: the main purpose of mobile edge computing (MEC) in cellular networks is to deploy computational, storage, and process capacity cloud resources on the edge within radio access network (RAN) to provide UEs with computing capabilities at the edge (BSs, Access points) of cellular network instead of relying only on centralized mobile cloud computing (MCC) [63]. Fig. 8 depicts the general architecture of mobile edge computing [64].

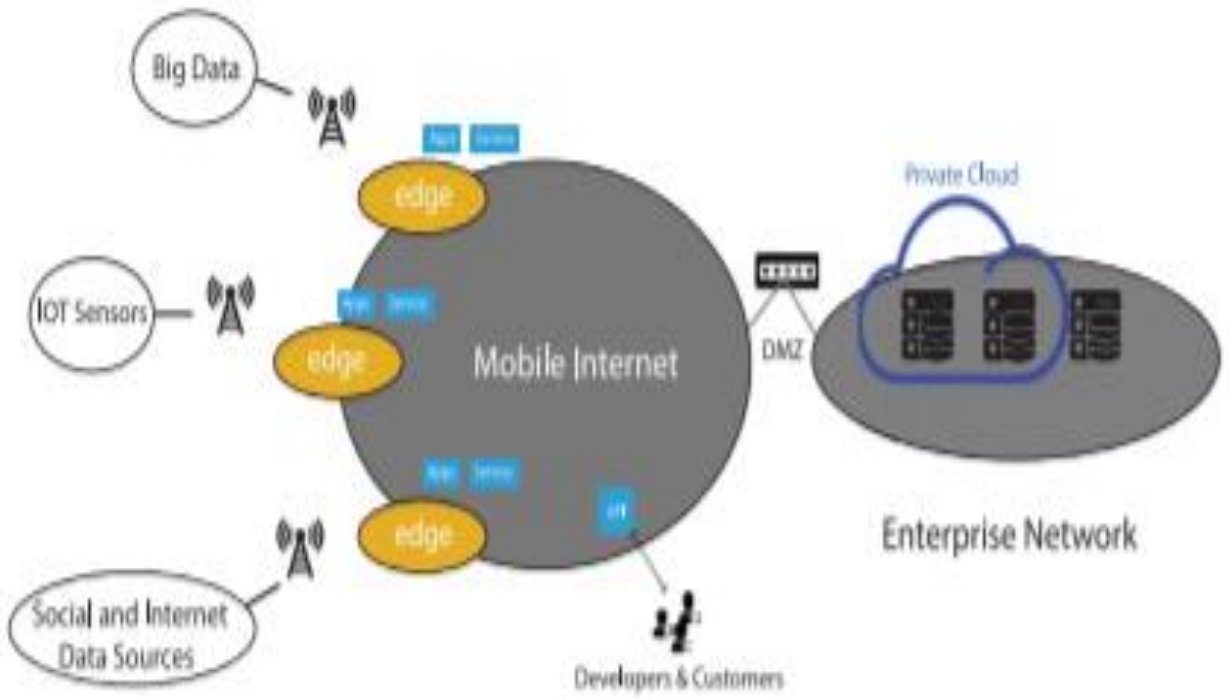

Fig. 8. Mobile Edge Computing MEC Architecture

Having computation resources at the edge of cellular network in 5G will bring benefits of offloading the backhaul transmission links, reducing latency as MCC cannot provide latency requirements for applications that require small latency values (< 1 ms), enhancing energy efficiency of user equipment and prolonging battery life of IoT devices [65].

## V. Resource Management In 5G

Resource management has been approached from different perspectives using models that can be utilized to achieve the best optimized results in wireless networks. These models include game theory, optimization and heuristic models, AI/machine learning. Resource management scheme selection is based on optimization objective like latency, spectral efficiency, energy efficiency, or throughput.

A. *Game Theoretic Mechanisms*:

Game theory is an applied mathematics technique used to model situations where decision makers or gamers must take actions that have mutual or conflicting consequences. In wireless, decision makers in the game are network users or operators controlling their devices to cope the limited resources (i.e., available spectrum), that can cause a conflict of interest [66]. In 5G, network base stations are considered as players that interact and can make decisions based on a set of actions that affect other base stations decisions. There are many papers that used game theory in 5G wireless resource management.

- *Evolutionary Game Theory (EGT)*:

  In [67], Inter-cell interference problem was addressed in scenarios when a layer of multiple femtocells sharing same frequency with an overlaid macro cell or macro base station (MBS). Stackelberg game was used where the control of interference from femtocell users is done by the MBS that prices UL interference from femtocell users based on allowable interference from MBS so that maximum utility of both femtocells users and macro cells users can be reached. The proposed Stackelberg model was implemented on rural femtocell network where interference between femtocells can be neglected, and dense femtocell network model where interference between femtocells is considered.

  As per [68], the authors proposed a distributed resource allocation scheme for DL transmission of OFDMA-SCs network underlying MBS using evolutionary game theory (EGT) to achieve evolutionary equilibrium for the players (in this case SBS) in self-organizing network. They proposed that all SCs choose transmission alignment randomly and then each player observes the received utility and sends it to system controller that calculates average system utility. This repeats at each iteration and each player compares the utility they receive with average overall utility of the system. If the received payoff is less than the average system utility, the player adopts new strategy aiming to achieve evolutionary equilibrium which represents the best scenario to manage and allocate resources over the players.

  In [69], the authors addressed the co-channel interference challenge resulted from adopting NOMA protocols in wireless networks where multiple users can transmit over the same subcarrier (power domain PD-NOMA) and the signal of one user can spread over multiple subcarriers (code domain CD-NOMA). They suggested that the adoption of evolutionary game theory (EGT) can overcome the complexity characteristics of NOMA compared to OMA. The implementation of game theory can be described as each user is a player who can switch from one mode to another according to the higher payoff the user can get when user utility falls below system average utility.

In a recent paper [70], two-layer game-based resource allocation model in an operator that offers two different levels of services with different QoS was adopted in integrated terrestrial-satellite network to maximizes operator payoff where resources are pooled in the cloud that acts as control center and the downlink of the satellite and the uplink of the terrestrial transmit over shared spectrum causing interference between the two levels. Authors proposed a scheme that combines three stages in which the first stage decides the price of available services, then in second stage users select from available and priced services using evolutionary game model, and in the third stage, and based on collected information from first two stages, the operator decides the power and resource allocation strategy.

As per results observation of adopting EGT model, it showed that it is suitable to tackle resource allocation problem since it allows the players to take decisions individually to reach equilibrium and allow fairness among players. But on the other hand, GMT are probably not able to model uncertainty nature of parameters (i.e., queue dynamics) and they assume the homogeneity of players, and they model the interaction of each player with other players so that they are not able to model large scale systems like 5G [71].

- *Mean Field Game (MFG)*:

  Unlike EGT, MFG models the interaction of a player with the overall impact of the behavior of all players in the system (mean field), making it better to model large scale networks [72].

  In [73], authors studied the problem of cache control and distributed caching in ultra-dense small cell networks (SCN), and they used MFG to address this problem. where each base station has storage to reduce the load on backhaul links by allowing users to access video contents cached in small base station SBS. The proposed model suggests that video file can be split into segments and each SBS can decide the size of segment it can store by considering storage state of other SBSs using MFG model, and the user can attain the file stored in multiple base stations. This model enhances storage space efficiency.

  In [74], authors also addressed computational caching problem in high dense cellular network. MFG was used to aggregate the behavior of SBSs in the network into a single interaction instead of gathering the information from each SBS directly. Then, each SBS can make its own caching decision. Thus, reducing the computation complexity of the model regardless of the number of base stations in the network. It also minimized the overlap of replicated content cached at the edge of the network that represents a waste of cache resources.

  MFG is appropriate for ultra-dense cellular networks and it helps reduce the load of transmission links and it also enhances system latency.

- *Minority Games (MG)*:

  Players' interactions in MG shows MFG-like performance in dense cellular networks and models congested systems with large number of players without the need for pair wise communication between players. In MG, players should be anonymous with random rationality, and their numbers must be odd and try to be in the minority for payoff. Future actions in this model are based on predefined set of strategies (that each player randomly selects from strategy e information from space) used to predict future wining action in the coming rounds upon analyzing previous winning actions [75].

  MG is used widely in cognitive radio networks (CRN) for efficient spectrum sensing where secondary users (SUs) can exploit the licensed spectrum that is not utilized by primary users (PUs).

  In [76], authors tackled the problem of existing spectrum sensing models encountered by CR environment due to accuracy issues, and proposed a novel model for better spectrum efficiency based on MG. In their model, players are SUs who can choose between to sense or not to sense the spectrum.

  Authors in [77] designed MG-based resource allocation so that CRs can allocate channels by playing minority game that minimizes information exchange between network players in the system to reduce traffic and relieve congestion.

One advantage of game-theoretic mechanisms is their ability to adapt to network dynamics and their flexibility, but these advantages are at the cost of high energy and computational requirements in each node so that in dense network environment, this is considered a waste of resources, in addition to the overhead generated by information updates among nodes. The other thing is that delay increases with the increase of network size and complexity.

B. *heuristic mechanisms*:

Different Heuristic-based techniques adopted in research to address resource management problems in cellular networks (i.e., network slicing, user clustering, radio block allocation…etc). These techniques include (but not limited to) heuristic, greedy, and metaheuristic methods.

- *Heuristic*:
  Heuristic models usually find good solution for a specific problem in short time but not global optimum solution. They may tend to adopt local optimum solution that could be far from achieving the optimum network performance.
  in [78], the authors proposed a heuristic-based resource management mechanism to enhance user experience through improving the delay of the system. When a set of users send requests for services in a network slice, the proposed heuristic mechanism decides the best server the slice should be assigned to, based on available CPU resources in the server, so that users' tasks processing time reduced and delay improves.
  Authors in [79] proposed a resource management and admission control algorithm based on heuristic model to dynamically allocate physical resources to each defined network slice to improve user experience, taking inter and intra slice priority into model's consideration. In this paper, the authors simulated their model for limited number of UEs and their proposed showed better data rates.
  In [80], the paper addressed the problem of network slicing when there are many simultaneous user requests need to be processed by the network. Authors used heuristic model to find the best route of traffic quickly to minimize network overhead. The paper does not consider user perspective and QoS upon using this model.
  In paper [81], authors used heuristic model to find a convenient way to allocate physical resources among the defined network slices to meet user requirements based on 5g use cases, where each case has its specific requirements of latency and availability.
  The advantage of heuristic model is that it allows to find resource management problem solution in less time because it does not require to evaluate all network nodes
- *Greedy*:
  Greedy algorithms tend to divide the problem into stages, sort them, and find a local optimum solution for each stage based on the given problem constraint.
  Authors in [82] addressed the problem of minimizing average sum downloading delay of video on-demand streaming by using greedy algorithm to place popular content at network edge in heterogenous cellular networks in which backhaul capacity is limited and considered a bottleneck, and the access points (SCs) have quite high storage to cache video files. The authors showed that the proposed algorithm can guarantee 50% of optimal content placement at network edge during off-peak hours through backhaul links that refresh edge storage during non-peak hours.
  In [83], authors addressed content caching issues and used greedy model to decide what data content should be cached at network edge so that UEs can access without using BH links, and what content need not to be cached so that the proposed system achieves best performance with least congestion in limited transmission links. Authors adopted greedy algorithm to increase cache hit rate in the network by caching the popular content at each network node. Greedy algorithm calculates the file contents to be cached based on the miss rate and hit rate of user requests.
- *Metaheuristic*:
  Metaheuristic techniques are high level heuristic, were developed to tackle multiple problems and they can overcome the previous heuristic approaches of generating local optimum solutions so that they generate better approximate solutions with the consideration of network dynamics and within reasonable execution time.
  In [84], authors addressed the problem of improving resource utilization and management of the spectrum for IoT devices deployed in the service area of 5G SCs. Metaheuristic algorithm is used to solve the issue of having tasks arriving in the same time to BS from small cells for transmission to core network. The algorithm may change the priorities of arriving tasks so that overall spectrum allocation improves if it is better than relying on the initial solution of higher-weighted tasks.
  Another field where metaheuristics applied is for determination of required number of BSs and their locations. In [85], authors studied link budget for 5G network and used metaheuristic algorithm to decide on the minimum number of BSs that can provide required capacity and coverage, in addition to realistic cell plan that minimizes the cost of deploying 5G BSs.
  Heuristic mechanisms are exploited in different papers to solve the problem of resource management because they are simple, easy to implement, capable to manage resources in reasonable time, and their results are locally acceptable, as they do not need to collect information on all available node resources for decision making. The disadvantages of these mechanisms is that they can only be used for specific problem domain, they are unable to give global optimal solutions for large scale networks like 5G, where accurate decision must be made as much as possible to make sure that best resource utilization is adopted to avoid KPIs degradation, resource utilization issues, network congestion and bad user experience, especially for services that need high requirements.

### C. *Machine Learning for Intelligent Resource Management:*

The aforementioned algorithms used in the literature for 5G resource management have shown good results, but with the increase of network size, the deployment of new nodes, and the rise of UE and network devices, their complexity increases and their ability to cope with the dynamics of real wireless network becomes unguaranteed. For example, using traditional heuristic algorithms with sensors could be very expensive computationally. Also, game theory techniques cannot support the heterogeneity of network devices and they also generate intensive overhead due to information exchange among players. Moreover, the accuracy of resource allocation using these algorithms declines with the expansion of network size and users specially that they usually generate local optimal. In intensive real-time data environments, such disadvantages are not acceptable. For all these reasons, adoption of machine learning ML, deep learning DL, Q-Learning, and reinforcement learning RL for resource management is important to exploit the huge amount of data generated by the wide range of applications and services in the cellular network to address resource management and allocation issues by extracting and inducing hidden features from the massive network data.

The authors of [86] utilized the data of human behavior, interaction with 5G application, and mobility through big data analytics and incorporated this knowledge to build a dynamic and efficient channel resource allocation scheme. In this paper, a contextual multi-armed bandit (CMAB) theory was used to build on-line learning algorithm called Agnostic Latent Stochastic Contextual multi-armed Bandit (A-SCB) problem that allocates channel resources dynamically to fulfil user QoE requirements more than focusing on network QoS. In this paper, based on the fact that user mobility can be predicted, geographic information of UEs were grouped into clusters created dynamically based on channel state information (CSI) distributions using the proposed agnostic stochastic multi-user algorithm A-SCB(D) so that upon the mobility of user, CSI distributions change among clusters and channel resources requirements change accordingly with taking QoE requirements reflected by (D) factor. A-SCB proved better performance in terms of predicting the best CSI for dynamic channel allocation in cellular system with the consideration of user mobility and geographic location than other classic bandits' algorithms and without implementation of ML.

In [87], authors combined both DL and Q-Learning to propose deep Q-Learning DQL-based algorithm called Action Reward Optimization Deep Q-learning (ARODQ) to enhance band utilization especially when spectrum resources are limited to guarantee the requirements of new 5G services. The proposed DQL-based system for making decision on resource allocation in real-time, slices the wireless channel conditions to have more accurate decisions and more efficient resource allocation of the spectrum. The results of the proposed system showed that the spectrum utilization improved especially when there is high load on the spectrum, the spectrum is utilized efficiently and up to the best of existing capacity compared to other solutions.

Authors in [88] also adopted DQL framework in cognitive radio network (CRN) to benefit from its capability in dealing with complex computation and the uncertainty of spectrum environment in terms of sensing and aggregating available channels. SUs sense available spectrum dynamically, taking into consideration user's transmission requirements for higher bandwidth that usually cannot be met by utilizing spectrum holes only, but through allowing spectrum or carrier aggregation (CA) of using multiple discrete aggregated channels in the same TS to achieve wider bandwidth without interrupting PUs. In addition, DQN allows users to make decisions of spectrum sensing and allocation without having knowledge of the dynamics of the system so that SU successful transmissions increases. The decision accuracy of proposed system shows almost same high accuracy compared to improvident policy that has full and prior knowledge of spectrum dynamics. This means that the learning capability of DQN is high and accurate.

Another study on improving cellular network capacity of CR communication system was carried out in [89]. The authors proposed an enhanced hierarchical learning model for spectrum availability prediction called DCG that benefits from both DL convolutional neural network (CNN) and gated recurrent unit (GRU). CNN was exploited to extract spectral features and mine availability of spectrum in existing channels within one TS (local mining) and within multiple TSs (regional mining), while GRU was used to mine temporal spectrum features on the long term, making spectrum availability prediction for SUs more accurate due to the consideration of both spectral and temporal features of the spectrum. An enhanced DCGEDCG model was also proposed in this paper and used to select an available channel between two SUs in CR system over which they can communicate. Compared with other deep learning models, this solution outperforms them with higher accuracy, recall, and intersection/union IoU values.

GRU also adopted in [90]. The authors proposed an AI-based framework to improve resource management of IoT data at 5G cloud datacenters called improved gated recurrent unit with stragglers detection (IGRU-SD) to predict future requests for resources at the cloud data centers to achieve efficient resource management strategies especially that fog/edge layers would not be able to provide sufficient resources for millions of IoT devices. The proposed module analyses the characteristics of tasks at the datacenter and classifies them based on their resource requirements. It also predicts future resources needed for the tasks to manage them with efficient utilization to avoid resource overprovisioning at data centers and reduce cost and energy consumption for the tasks at the cloud. The authors compared between the predicted or future required resources using proposed model with other resource management models at data centers and they found that their proposed neural network-based model achieved the best prediction accuracy. Such results can be exploited by data centers on the long term to build strategies that result in shutting redundant servers down to save energy and enhance resource utilization.

ML techniques also used in V2V resource management. In [91], the authors addressed improving V2V communications and resource management. through adaptively changing transmission modes for communication as well as the power level. RL was exploited to select the optimal transmission mode and required level of power to maximize link capacity of vehicle to infrastructure (V2I) links and fulfil reliability and delay requirements for vehicle to vehicle (V2V) links. Double deep q-learning algorithm (DDQN) was used to design a frame work that instantaneously observes the dynamics of vehicular environment and learn how to adaptively allocate best capacity and power resources over V2V link which is considered as an agent that targets increasing the reward by learning the best power and transmission modes that improve V2V performance to generate more accurate prediction of controlling parameters and make the decisions intelligently. The results show that DDQN could jointly select transmission mode and adaptive power in V2V link under vehicular environment dynamics (channel fading of mobility) in short time. The authors measured the data throughput with the for moving vehicles using different learning techniques, and the proposed technique in this paper.

Authors of [92] addressed the services that require high reliability and low latency (uRLLC) because these services can be realized and improved by using ML for controlling the flow of mobile network traffic. In this paper, a prediction model for the traffic flow was developed based on long short-term memory (LSTM). LSTM deep learning algorithm is able to classify and predict long term dependency data, and it was exploited for uRLLC data flow prediction in single-site, so that edge-clouds utilize the algorithm to predict traffic flow and report the prediction to the remote cloud that analyzes the received reports and allocate resources intelligently to avoid sudden increase of traffic on network edge-cloud that may cause congestion and negative impact on uRLLC services. The authors proved that even with small training data for the model, it was able to make good prediction when compared to actual traffic flows. In addition, the model was able to noticeably reduce packet loss in network using deep learning model compared with the scenario where machine learning and prediction techniques were not used.

In [93], the authors address the autonomous of $5^{th}$ generation networks based on zero touch network and service management (ZSM). For this purpose, an Adapted REinforcement Learning VNF Performance Prediction module for Autonomous VNF Placement (AREL3P) was proposed to enhance the functionality of network orchestrator through on-line learning to manage VNF resources. The authors used Q-learning as a type of RL to discover unknown parameters and dynamics of the environment because it does not require complete prior knowledge, so that the proposed model can be adaptable and can be generalized over a heterogenous networks as a feasible and adaptable model. In addition, this model can forecast end-to-end service performance for more accurate decision making on placing VNF. The results showed efficient placement of VNF compared to other supervised models, and increased prediction accuracy as well.

In [94], the authors tried to build a model that can schedule users in a dense cellular network efficiently and reduce the impact of inter-cell interference resulted from the distributed method of managing scheduling in heterogenous network with spectral-temporal variations specially for real time services. The approach is designed in a way that a scheduling method can be selected dynamically from multiple scheduling methods by a central agent the manages the selection over a group of cells to achieve best experience for users of these cells. This was done by using RL for dynamic scheduler selection (RL-DSS) that learns from traffic load and channel quality indicator (CQI) reported by the users of each cell to predict the benefits or rewards for each action. DNN is also used to approximate functions in RL because the amount of data to be analyzed will be too large and conventional contextual bandit techniques are not able to scale such data properly. The performance of proposed algorithm was compared against genetic algorithms (GA) and convolutional algorithms. The results showed that RL-based and GA-based algorithms have better user satisfaction rates than convolutional algorithms. In addition, RL-based algorithms showed better performance than GA algorithms on the long term because it needs more time for training.

To improve energy resource management and enhancement of MEC system and reduce computation resources, [95] addressed the problem of improving video streaming in MEC-enabled software defined mobile network. The authors utilized DRL - asynchronous advantage actor-critic (A3C) with markov decision process (MDP) taking into consideration different factors of caching, streaming, and transcoding. In this case, MDP can track the true dynamics of the environment unlike other non-markov models. The simulation of proposed MDP with A3C showed better throughput and less time for video stalling, and on top of that, the energy required for video streaming services became less.

Below table demonstrates the latest papers in 2020 that addressed 5G cellular networks resource management based on ML techniques and blockchain technologies and the algorithms used to provide proposed solutions.

TABLE. II
Recent Intelligent & blockchain Resource Management Based on ML

| Ref | year | Proposed Model | Problem | Solution | ML Technique |
|---|---|---|---|---|---|
| [86] | 2020 | A-SCB(D) | Intelligent allocation of spectrum resources | Incorporate user mobility information in channel allocation decisions | DLP-A-SCB |
| [87] | 2020 | ARODQ | Bandwidth utilization enhancement in limited spectrum network | Slice channel conditions for better learning and accurate spectrum resource allocation | DQL |
| [88] | 2020 | DQN framework for POMDP | Improving SU spectrum sensing, aggregation, and successful transmission | Correlate between channels in wireless network and model the dynamic spectrum environment as a joint Markov chain | DQL |
| [89] | 2020 | DCG/EDCG | Spectrum availability prediction and enabling SUs communication on the same channel | Build hierarchical spectrum learning model to predict local spectrum availability for each SU without any prior information of Pus with taking into consideration both temporal and spectral features | DL/CNN |
| [90] | 2020 | IGRU-SD | Predict the level of resource requests at cloud data centers to avoid resource overprovisioning | Classify tasks based on their resource intensity and predict the expected level of resource requests, and allocate enough resources to handle the jobs | RNN |
| [91] | 2020 | DDQN framework | Improve the capacity of V2I links to guarantee QoS of V2V communication with fast changing wireless channels | Maximize V2I capacity and optimizes interference management and power adaptation based on different transmission modes to improve the performance of the V2V communication network | DL and QL |
| [92] | 2020 | LSTM | Control mobile traffic flow in 5G | train mobile-traffic data in single-site mode to predict peak traffic volume and build an architecture that enables remote cloud to collect traffic flow predictions of multiple sites | DL |
| [93] | 2020 | AREL3P | The autonomous placement of Virtual Network Functions (VNFs) | monitor and forecast e-2-e service-level performance across system layers spanning from the network layer up until the application layer | RL(QL) |
| [94] | 2020 | RL-DSS | Manage distributed scheduling functionality to reduce inter-cell interference | manage distributed scheduling methods across a small cluster of cells by dynamically selecting schedulers to be implemented at each cell | DRL |
| [95] | 2020 | A3C/MDP | Energy conservation for video streaming services over SDMN with MEC | MDP provided the ability for tracking true environment dynamics in addition to taking multiple network attributes into consideration | A3C - MDP |

### D. *Blockchain-based resource management:*

Blockchain has an increasing impact on handling wireless networks resources in smart way. We will cover few recent papers that jointly studied blockchain with future wireless networks and technologies:

In [96] [97], the authors addressed the problem of managing spectrum resources in dense wireless network with MEC system. The authors proposed decentralized blockchain-based MEC framework (B-MEC) for resource management and adaptive resource allocation in future wireless networks. The proposed framework suggests a consensus protocol based on (PBFT) and (DPoS) and is described as markov decision process (MDP) that is handled with DRL to solve the problem of highly dynamic system. This scheme was compared with other schemes for evaluation. B-MEC model has achieved the highest rewards in all the experiments, reflecting the high potential of using blockchain with resource management.

The authors in [98] also addressed the importance of using blockchain with wireless networks like 5G. The authors studied the problem of minimizing energy consumption in MEC system and reducing delay time to finality (DTF) of blockchain system in a blockchain-based MEC scenario. Results showed that the proposed system achieved the best trade-off between MEC energy consumption and DTF.

In [99], the authors employed blockchain to build a decentralized scheme for resource allocation that was able to reduce latency and improve the speed of system provisioning and computing capabilities in an IoT with MEC system.

## VI. Quality of Service Provisioning Schemes In 5G

In previous sections, we covered the main services and applications promised by 5G and B5G networks. We also covered recent techniques and schemes for resource management that will have important role in 5G networks. Another important topic that is also critical to guarantee stable and efficient performance is to provide quality of service (QoS) guarantees to the new 5G applications and services, since QoS is one of the main objectives that need to be carefully addressed when designing 5G technologies. The diversity of applications supported by 5G requires diversity of QoS requirements to cope with the needs for multiple data rates, low latency values, high reliability, low delay, and low packet loss. QoS is provided based on different network components such as admission control (AC), routing, scheduling, resource scheduling, and interference management [100].

In this section, we are going to survey QoS provisioning schemes in 5G cellular networks and their contribution in rolling out the wide range of services that 5G is promising by guaranteeing required network resources in a varying cellular load.

### A. *Admission Control (AC)*:

*Call Admission Control (CAC)* was designed to control the acceptance of both new calls and handoff requests with specific QoS requirements and to also maintain QoS guarantees to existing calls whether they real time (RT) or non-real time (NRT). In other words, CAC aims to achieve efficient allocation of network resources and keep monitoring resource utilization when network resources are overloaded to avoid congestion. CAC process in wireless technologies admits or rejects new UE requests to connect to network based on different parameters including availability of resources, quality policies, priority of the call, mobility, and network optimization strategies. We will cover recent research on improving CAC schemes in 5G cellular networks and their role in avoiding network congestion.

In [101], the authors addressed the admission control for coexisting uRLLC (low reliability/high reliability requirements) – eMBB (high data rates requirements) users. Based on services' requirements, all uRLLC users should be prioritized over eMBB users, and the later should controlled by admission control. The proposed scheme aims at maximizing the number of eMBB users admitted in the cellular network with guaranteeing their data rates, while all uRLLC users are scheduled for required resources allocation to meet Qos requirements, taking signal-to-interference-plus-noise (SINR) ratio for eMBB users, signal-to-noise ratio (SNR) for uRLLC users, transmit power, and bandwidth constraints. The results showed that the algorithm optimized the bandwidth utilization and it was able to admit the maximum eMMB users with guaranteed data rates while facilitating the allocation of power and bandwidth to all uRLLC users.

In [102], the authors emphasized on the importance of adopting new CAC schemes instead of traditional schemes in C-RAN to maintain the QoS of RT and NRT services in 5G cellular network. BSs' baseband resources are shred in a BBU pool in the C-RAN and can be used in multiple cell areas so that they can cope with traffic variations in the network. A new CAC scheme in 5G C-RAN was proposed based on fuzzy logic that helps in making accurate decisions from approximate information to avoid 5G network congestion resulted from billions of connection requests that need to be processed by the C-RAN. In addition, the authors proposed to preempt NRT connections to be processed in public C-RAN so that 5G network C-RAN processes high priority real-time traffic that requires guaranteed QoS requirements. Fuzzy controller is used in this scheme to make CAC decisions of admission/rejection/preemption for new connection requests. The authors conducted a comparison to verify the results and they found that CAC decisions in C-RAN with fuzzy logic outperforms the results of traditional admission schemes. Blocking and dropping probability reduced noticeably in fuzzy logic C-RAN and overall system throughput improved.

AC is also used to grant admission to tenants of each virtual slice as per QoS requirements and service level agreement (SLA) for the service that the slice will provide resources to. We will go through different AC mechanisms on network logical slices and their contribution in improving network performance.

The authors in [103] stressed on the importance of picking the most appropriate network slice that fulfil the QoS requirements for the type of connection accessing the network. The slice orchestrator can be supported by appropriate ML technique that can efficiently predict the perfect slice for a new service based on learning from historical data. The authors carried out predictions using different ML models (KNN, SVM, Random Forest, Decision Tree) and their work resulted in an accuracy of around 99.89% (KNN classifier) in picking the best slice for a given service.

In [104], slice admission control based on reinforcement learning was proposed. The paper aimed to use RL to maximize InP revenue by predicting whether to admit or reject slice connection request based on latency requirements that defines its priority (high priority HP or low priority LP) so that the lower the latency for the service, the higher the priority and the more revenue it brings to the infrastructure provider InP. Failure to admit a connection request results in a penalty on the InP resulting in a loss of the revenue. To improve admission performance, RL was used with AC to help make decisions on scaling up/down the slices to redistribute resources among them and guarantee that the requirements of high priority services can be fulfilled. The results showed that using RL with AC achieved reduction of more than 54% on the penalty amount paid by the InP due to service rejections.

In another paper [105], the authors proposed slice AC scheme called MSRAA for inter-slice resource allocation based on two algorithms. The first one is bandwidth AC (BAC) that utilizes LSTM neural networks to predict bandwidth requirements and guarantee bandwidth QoS for high priority services. The second algorithm called delay AC (DAC) which adopts mandrian random forest to predict e-2-e delay and admits services based on delay requirements. The proposed scheme assumes the existence of multiple slices handling traffic of different priorities so that GBR eMBB generates more revenue and has higher priority over N-GBR services. The scheme also uses machine learning techniques to forecast resources for services that are both bandwidth and delay guaranteed. Forecasting bandwidth and delay helps to increase acceptance of GBR connections and improve resource utilization with limited impact on N-GBR connections. This scheme increases network profitability through increasing the admission of GBR flows.

The authors of [106] stressed on the importance of proper admission of slice requests on users' QoS and the impact of network traffic dynamics on resource allocation. The authors studied these factors in a limited resources MEC system to increase the average revenue of an operator with the consideration of traffic variations. For this purpose, a framework to maximize operator's revenue with the assistance of a proposed algorithm called Dynamic Network Slicing and Resource Allocation (DNSRA) that helps optimize dynamic slice admission. The proposed algorithm outperformed other algorithms handling MEC limited resources as the proposed algorithm showed more revenue with limited resources.

In [107], the authors proposed enhanced V2X (eV2X) communication by using improved QoS model called multi-level QoS (MLQ) that mainly targets safety services for V2X communications. MLQ can improve the QoS for V2X when there are bad conditions on the access node like congestion and interference. In case of performance degradation and the required QoS attributes cannot be met, the traditional approach to handle this situation is to downgrade the QoS model to *best effort (BE)*, that may result in a service interruption. The proposed alternative MLQ approach is designed to define preferred QoS profile in addition to a set of acceptable alternatives that have lower QoS requirements. With MLQ, even with bad conditions, the connection will not transfer to BE, otherwise, it will adapt to the best QoS profile in the predefined set that increases the chances of maintaining the service. Results showed that MLQ minimized service interruptions for V2X by reducing session rejection rate and handover failure rate by about 36%.

### B. *Routing*:

Routing is an important component of QoS, and it has an important contribution to improve services in 5G networks. We will cover some of recent papers related to routing in 5G networks.

Load balancing is a core function that self-optimized networks (SON) in 5G need to fulfil to avoid performance degradation due to cell overload. In [108], the authors used QL techniques to improve mobility load balancing (MLB) so that the process of redirecting traffic from congested cells to other cells would be more efficient. This paper proposes an algorithm called user specific, optimal capacity and shortest path (US-OCSP) to dynamically route traffic from overloaded cell to ideal cell in terms of the shortest possible path and available capacity. Capacity in this paper is measured based on the availability of physical resource blocks (PRB) for the cells. Using Q-Learning helps determine the least distant cell that has enough resources to serve end user connection so that throughput and QoS requirements can be met and network connection rejections can be reduced. The results of implementing US-OCSP showed that the algorithm was able to find the shortest path node that has the optimal available resources for load balancing.

In [109], the authors addressed the problem of using new source routing technique called *Segment Routing "SR"* so that an SDN controller can compute packets path to destination, and SR segments the overall computed path into segments representing the links/nodes to the destination. Segment identifiers SIDs are encoded as a list of labels and stored in the packet. Thus, SR adds overhead to the packet resulting in more bandwidth requirements and higher delay. This paper proposed a novel routing technique called preferred path routing (PPR) to avoid the list of labels that should be added to the packet. Instead, PPR provides the advantage of adding one or two labels only to the packet to identify the routing path. Results showed that the overhead significantly minimized leading to better QoS and lower delay.

Authors in [110] focused on improving enhanced applications such as virtual reality and video gaming and streaming through i) splitting large multimedia flow into smaller flows transferred to the destination over separate network paths decided by a central controller, where each flow is controlled by the congestion control mechanism to make sure that available resources are enough on any allotted path by using multi-path transmission control protocol (MPTCP), and ii) using advanced packet forwarding in 5G called *segment routing (SR)*. The proposed scheme is called MPTCP/SR, where the utilization of SR removes the limitation of extra load on SDN switches due to using MPTCP. SR encodes routing information as ordered labels into the header of packets. This significantly reduces forwarding rules in SDN switches, reduces the need for extra expensive memory resources in SDN switches, and improves QoE for multimedia services. Authors also proposed quality of experience centric multipath routing algorithm (QoMRA) to control the number of sub-flows derived from the main flow dynamically and to route the sub-flows by using SR to a new path if the path allocated by MPTCP failed due to not enough resources. the proposed scheme and algorithm achieved much better throughout compared to traditional TCP routing, and significantly improved link utilization, and improved QoS/QoE.

In [111], the authors addressed the problem of the huge overhead of rule updates resulted from user mobility and handover. The paper suggests using segment routing SR to reduce the negative impact of rule update overhead so that the network provides efficient live video streaming for 5g network users. The proposed algorithm was called mobility aware multicast tree algorithm (MAMTA) that predicts the BS that can serve the connection for the longest possible time to avoid handover and rule updates in SDN switches as much as possible. MAMTA builds a multicast tree and installs the rules on each router on the path. Results showed that MAMTA has better resource utilization and QoS, compared to other tree algorithms especially in MAMTA adaptation to 5G network dynamics and in its lower cost.

In [112], the authors focus on efficient resource allocation in SDN-based 5G network with the existence of multiple paths forwarding to connect users with servers. Their approach is based on calculating routing paths for existing users, new users, and users in mobility periodically to always find the best route for users with taking network dynamics and QoS requirements into consideration to achieve the best resource allocation. The authors compared two techniques for this purpose, i) Heuristic paths re-computation (HPR) that considers paths reconfiguration for all network users, and , ii) partial paths re-computation, that avoids path reconfiguration for constant users who have same location and same QoS requirements (new users or HO users). The authors studied the impact of adopting proposed algorithms on network reconfiguration and found that PPR consumes less resources for computation than HPR. Moreover, HPR would have less computation time than PPR but with much higher consumption of resources.

*New Ip* network protocol introduced in [113] to overcome the deficiencies of traditional IP protocol that will not be able to fulfil the requirements of 5G and B5G due to its fixed structure that prevents flexible changes in services, payloads, and addresses. New IP provides flexible addressing of network elements and supports service/application – awareness to guarantee QoS requirements of reliability, latency, and capacity; and it also improves throughput and data robustness for the new services emerging in 5G networks.

C. *Resource Scheduling*:

Authors of [114] addressed the problem of overlapping eMBB transmissions by uRLLC traffic causing QoS degradation in eMBB services and thus the loss of eMBB revenue that could be more than the revenue of uRLLC services. The authors used DRL to learn the optimal performance tradeoff between uRLLC/eMBB transmissions and they proposed a method called system-wide tradeoff scheduling (SWTS) and compared it with user-centric tradeoff scheduling (UCTS). Proposed SWTS method achieved the best eMBB QoS and reliability with slight degradation of uRLLC services, while UCTS achieved the best uRLLC performance at the cost of huge eMBB QoS degradation. In other words, SWTS guaranteed better QoS for more eMBB users and it also achieved the best tradeoff performance.

In [115], the authors presented QoS aware scheduling scheme that supports GBR and NGBR data radio bearers (DRB) jointly for eMBB services called enhanced joint scheduling (eJS), with guaranteeing minimum QoS requirements for GBR services. The proposed scheme can support two different strategies, the first one called best CQI highest deviation (BestCQI_HD) that boosts system throughput at the expense of fairness among DRBs, while the second one is called best CQI lowest second (BestCQI_LS) that enhances DRBs fairness with slight system throughput degradation. Both strategies consider the impact of scheduling a DRB on system throughput and fairness. The proposed scheme was compared with different existing schemes for resources scheduling and the results showed that eJS outperforms other scheduling techniques through achieving the best tradeoff between system throughput and resource allocation fairness for better overall system QoS.

Authors in [116] focused on the importance of resource scheduling to improve the gain of resource multiplexing among network slices so that RAN resources are shared adaptively and flexibly among tenants of different slices with maintaining a certain level of isolation among slices to fulfil QoS requirements in each slice. To achieve this goal, the authors proposed intelligent resource scheduling system (iRSS) for resource sharing among RAN slices. The system was presented in a learning framework where both DL and RL were used to work collaboratively. The former exploited by using LSTM for large time scale prediction of RAN slices resources, while the latter uses asynchronous advantage actor-critic (A3C) algorithm to perform on line RAN slices resource scheduling to cope with inaccurate prediction in short time scale that DL (LSTM) cannot satisfy. The proposed frame for resource scheduling of RAN slices outperformed other traditional and intelligent resource scheduling techniques in term of resource utilization.

In [117], the authors proposed two level-scheduler for heterogenous mobile network slices, where each slice has its own QoS requirements. The first level scheduler is the *global scheduler* that is responsible for initiating the second scheduler called local *slice scheduler* for each network slice and allocate resources and cluster of nodes to it as per local slice needs and QoS requirements. On the other hand, the slice scheduler locally controls and analyzes the allocated resource to the slice through an algorithm that is set up and configured by the global scheduler. Both global and slice schedulers share information about allocated and available resources of the nodes belonging to the slices to facilitate resource control operations. The proposed multi-level scheduler was compared with one level scheduler in term of establishment time, and the results showed that the time required to deploy multi-level scheduler shortens and becomes less than the time needed in one level scheduler as the number of created slices increases.

In [118], the problem of scheduling MTC class of devices was addressed. Traditionally, MTC devices are provided with preference by using SNR only, thus ignoring different QoS requirements of the MTC class of devices and providing priority to MTC devices that are located close to the MTC gateway to send transmission requests. Moreover, the fact that some MTC applications dedicated to accomplishing one task may result in an outage because not all MTC devices in the class were able to send their transmissions due to the adoption of conventional random-access techniques. The authors proposed two-phases scheduling algorithm medium access phase and allocation phase. At medium access phase, devices priorities are calculated based on latency requirements in the MTC class, waiting time for transmission, and the received SNR. The second phase called allocation phase, allocates resources to MTC devices based on the calculated priority. The proposed scheduling scheme showed that outage probabilities significantly reduced with the proposed scheme and probabilities of successful MTC transmissions also increased.

In [119], the authors addressed the challenge of offering a suitable scheduling scheme to fulfil user-level QoS for real-time and non-real time services. A resource scheduling scheme for user level quality of services QoS requirements fulfilment for orthogonal-FDMA system with carrier aggregation CA called algorithm is called improved joint user carrier scheduling (IJUCS) was designed to guarantee user level QoS demands and fair allocation of radio resources by introducing service weight factor that changes adaptively in each TTI to reflect the proportion of RB that each user can utilize. The performance of improved joint user carrier scheduling IJUCS was compared with the traditional joint user scheduling JUCS and the results showed that IJUCS has better minimum throughput for non-real time (N-RT) users, less delay for real-time RT users, and much less packet drop rate PDR.

### D. *Interference Management:*

Interference in 5G heterogenous networks has negative impact on QoS. Managing interference has become critical to maintain good connections and meet QoS requirements. In [120], the authors addressed the importance of interference management in ultra-dense network (UDN) that is based on NOMA technology. In such scenario, spectrum is shared universally by SBSs and users causing increase in interference. Motivated by reducing inter-cell interference of overlapped coverage SBSs and intra-cell interference of NOMA technology, the authors designed an efficient mechanism to reduce interference in such networks called self-optimizing interference management (SOIM) in a PD-NOMA based UDN network. each SBS generates interference graph including available resources and SOIM adaptably performs user and sub-band selection, and power allocation. The results showed that SOIM has better interference management performance compared to other schemes and it also showed improvement in spectrum efficiency and throughput.

TABLE III
5G QoS Schemes

| Ref | **Year** | **Schemes/Techniques for QoS** | **Problem** | **Technique/Algorithm** | **Constraints** | **Results** |
|---|---|---|---|---|---|---|
| [101] | 2020 | CAC | uRLLC/eMBB Admission Control | Approximation Methods/sequential convex programming | • eMBB SINR<br>• uRLLC SNR<br>• Power<br>• Bandwidth | Bandwidth allocated optimally so that more eMBB users can be admitted while allocating resources to all uRLLC users. |
| [102] | 2017 | CAC | Admission/Congestion Control in 5G C-RAN | Fuzzy logic | • Effective Capacity<br>• Service Type<br>• Available Capacity | Blocking/dropping probability reduced, and network throughput increased |
| [103] | 2019 | Slice AC | Allocating best slice for a given service | ML<br>• Random Forest<br>• SVM<br>• KNN<br>• Decision tree | • Throughput<br>• Latency<br>• Reliability | 99.89% accuracy in selecting network slice |
| [104] | 2018 | Slice AC | Increase InP revenue through slice AC scheme for connection requests | RL | • HP/LP slices<br>• Latency<br>• InP Profit | Efficient resource utilization and around 54% less penalty by InP, resulting in higher revenue |
| [105] | 2020 | Slice AC | Inter-slice AC | ML<br>• LSTM<br>• Mondrian random forest | • GBR eMBB/N-GBR eMBB<br>• Bandwidth<br>• Delay | Increase GBR eMBB admitted connections and overall system profits |
| [106] | 2020 | Slice AC | Increase operator's revenue with the consideration of traffic variation | DNSRA | • Operator's revenue<br>• Traffic variation | Improved operator's revenue in different scenarios compared with other schemes in the literature |
| [107] | 2020 | V2X AC | Reduce V2X service Interruptions | MLQ | • Throughput<br>• Latency<br>• Reliability | Session rejection rate/HO failure rate were reduced with acceptable QoS using multi-level QoS |
| [108] | 2019 | Routing | Load Balancing in SON networks | Q-Learning | • Physical Resource Block PRBs<br>• Shortest path | Fast determination of optimal nodes for traffic redirection |
| [109] | 2018 | Routing | Segment Routing SR overhead | PPR | • Bandwidth<br>• Delay | Overhead minimization, Less bandwidth and lower delay |
| [110] | 2019 | Routing | Multimedia traffic routing | • MPTCP/Segment Routing (SR)<br>• QoMRA | • QoE/QoS<br>• Link utilization<br>• Data rate | Higher data rates, better link utilization, and improved QoE |
| [111] | 2018 | Routing | Live video multicast | • MAMTA | • QoS<br>• rule update overhead | Less rule update overhead achieved and more efficient routing tree |
| [112] | 2020 | Routing | Routing reconfiguration | • HPR<br>• PPR | • Computation time<br>• Resource Consumption (OVS) | Trade off between re-computation time and resource consumption between the proposed routing algorithms |
| [113] | 2020 | Routing | Traditional IP deficiencies | New IP protocol | Maintain fundamental internet architecture | - |
| [114] | 2020 | Scheduling | Improving trade-off between eMBB/uRLLC | SWTS | eMBB QoS | Better QoS for more eMBB users and better trade-off performance. |
| [115] | 2020 | Scheduling | Improving throughput/fairness trade-off between GBR/NGBR eMBB services | eJS<br>• BestCQI_HD<br>• BestCQI_LS | • PRB resource allocation<br>• System throughput/fairness | Better trade off between throughput/fairness compared with other techniques |
| [116] | 2019 | Scheduling | Improving resource multiplexing gain among network slices | iRSS<br>• DL: LSTM<br>• RL: A3C | Resource block utilization | The proposed framework (iRSS) achieved better resource utilization |
| [117] | 2019 | Scheduling | Multi-level network slices resource scheduling | Global scheduler/local slice scheduler | Time of deployment compared to one level scheduler | Multi-level scheduler deployment time reduces as the number of slices increases to be equal or even less than the time needed for one-level slice scheduler |
| [118] | 2020 | Scheduling | MTC application outage and MTC data transmission success | Improved scheduler with preferential access | • MTC delay requirements<br>• SNR<br>• MTC transmission request waiting time | Reduced MTC application probability of outage and improved MTC data transmission success rate |
| [119] | 2018 | Scheduling | User level QoS aware resource scheduling for RT/NRT in OFDMA system | IJUCS | • PDR<br>• RT Delay<br>• NRT min throughput | IJUCS showed better min throughput for NRT user traffic, less delay for RT users, and less PDR for RT services than JUCS |
| [120] | 2020 | Interference Management | Mitigate interference in NOMA based UDN | SOIM | • Inter/intra cell Interference<br>• Throughput<br>• SE | SOIM was able to reduce interference and meet QoS requirements of users. Better SE and improved outage probability |

## VII. Conclusion

In this paper, we covered the recent papers that addressed resource management schemes and QoS provisioning schemes in 5G cellular networks. In section III, we started with a review of 5G networks specifications that 5G networks promise to offer like latency, throughput, reliability, spectrum efficiency and energy efficiency as per TABLE I, Then we covered some technologies and enablers of 5G cellular network that will contribute in meeting high 5G networks requirements like mmWave, beamforming, D2D, and IoT. Then in section IV, we covered some of the networking techniques that allow the shift toward 5G like virtualization, softwarization, programmability, edge computing, network caching. In Section V, we surveyed the latest research papers that addressed resource management schemes in 5G cellular networks, and we found that papers have utilized different algorithms and mechanisms to address this critical topic, like Game Theoretic Mechanisms, heuristic mechanisms, Machine Learning. As per our comparative analysis for these papers we found that machine learning and more specifically reinforcement learning techniques, outperform other methodologies because they need less computation resources, they are appropriate for heterogenous networks with huge number of nodes and links, and because of their ability to learn from the environment and predict the required resources to achieve the best possible resource utilization and meet 5G network requirements even with limited resources. Most recent papers in 2020 studied managing 5G network resources using ML and NN as per TABLE II.

Then in section VI, we covered how 5G networks will be able to meet the quality of service requirements of the wide range of services that 5G networks support, and we reviewed recent papers and contributions on QoS provisioning schemes. We surveyed some of the components that contribute in fulfilling QoS requirements like admission control, routing, data scheduling, and interference management, and the algorithms and techniques utilized to improve the performance of 5G networks to meet QoS needs. In TABLE III, we listed the most recent papers, their techniques, and results. We can say that AI and ML have showed promising results.

In future, more focus on using big data analytics and deep learning should be given for improving resource management and QoS provisioning intelligently. AI and all the algorithms under this field have shown outstanding performance in real networks because of their compatibility with 5G networks that are flexible, scalable, and reliable.

## Reference

[1] V. J. Reddi, H. Yoon, and A. Knies, “Two Billion Devices and

Counting," IEEE Micro, vol. 38, no. 1, pp. 6–21, 2018.

[2] H. Shah-Mansouri, V. W. Wong, and J. Huang, "An Incentive Framework for Mobile Data Offloading Market Under Price Competition," IEEE Transactions on Mobile Computing, vol. 16, no. 11, pp. 2983-2999, Jan. 2ab017.

[3] "Cisco Visual Networking Index: Global Mobile Data Traffic ..." [Online]. Available: https://s3.amazonaws.com/media.mediapost.com/uploads/CiscoForecast.pdf

[4] Oliver, Nuria, Emmanuel, Harald, D. Nadai, Marco, Bruno, Lambiotte, Renaud, Richard, Ciro, de Cordes, Nicolas, S. P., Lehmann, Sune, Murillo, Juan, Alex, Pham, P. N, Salah, A. Ali, Saramäki, Jari, Scarpino, S. V., Michele, Verhulst, Stefaan, Vinck, Patrick, and Benjamins, "Mobile phone data and COVID-19: Missing an opportunity?," arXiv.org, 27-Mar-2020. [Online]. Available: https://arxiv.org/abs/2003.12347

[5] Q.-V. Pham, D. C. Nguyen, T. Huynh-The, W.-J. Hwang, and P. N. Pathirana, "Artificial Intelligence (AI) and Big Data for Coronavirus (COVID-19) Pandemic: A Survey on the State-of-the-Arts," 2020.

[6] G. Song, W. Wang, D. Chen, and T. Jiang, "KPI/KQI-Driven Coordinated Multipoint in 5G: Measurements, Field Trials, and Technical Solutions," IEEE Wireless Communications, vol. 25, no. 5, pp. 23-29, 2018.

[7] J. Zhang, Y. Xiao, D. Song, L. Bai, and Y. Ji, "Joint Wavelength, Antenna, and Radio Resource Block Allocation for Massive MIMO Enabled Beamforming in a TWDM-PON Based Fronthaul," Journal of Lightwave Technology, vol. 37, no. 4, pp. 1396–1407, 2019.

[8] J. G. Andrews et al., "What Will 5G Be?," in IEEE Journal on Selected Areas in Communications, vol. 32, no. 6, pp. 1065-1082, June 2014, doi: 10.1109/JSAC.2014.2328098.

[9] B. Lindoff, "Using a direct conversion receiver in EDGE terminals-a new DC offset compensation algorithm," 11th IEEE International Symposium on Personal Indoor and Mobile Radio Communications. PIMRC 2000. Proceedings (Cat. No.00TH8525), London, UK, 2000, pp. 959-963 vol.2, doi: 10.1109/PIMRC.2000.881564.

[10] A. R. Mishra, "Advanced Cellular Network Planning and Optimisation: 2G/2.5G/3G...Evolution to 4G," Wiley.com, 11-Jan-2007. [Online]. Available: https://www.wiley.com/en-us/Advanced Cellular Network Planning and Optimisation: 2G 2 5G 3G Evolution to 4G-p-9780470057636

[11] S. I. Shah, "UMTS: High Speed Packet Access (HSPA) Technology," 2008 IEEE International Networking and Communications Conference, Lahore, 2008, pp. 2-2, doi: 10.1109/INCC.2008.4562669.

[12] "Opportunities in 5G Networks: A Research and Development Perspective," CRC Press. [Online]. Available: https://www.routledge.com/Opportunities-in-5G-Networks-A-Research-and-Development-Perspective/Hu/p/book/9781498739542.

[13] A. Büyükoğlu, M. İ. Sağlam, A. Kavas and M. Kartal, "An efficient throughput averaging method for proportional fair algorithm used in mobile networks," 2016 Advances in Wireless and Optical Communications (RTUWO), Riga, 2016, pp. 161-166, doi: 10.1109/RTUWO.2016.7821876.

[14] K. Zheng, B. Fan, J. Liu, Y. Lin and W. Wang, "Interference coordination for OFDM-based multihop LTE-advanced networks," in IEEE Wireless Communications, vol. 18, no. 1, pp. 54-63, February 2011, doi: 10.1109/MWC.2011.5714026.

[15] B. Bangerter, S. Talwar, R. Arefi and K. Stewart, "Networks and devices for the 5G era," in IEEE Communications Magazine, vol. 52, no. 2, pp. 90-96, February 2014, doi: 10.1109/MCOM.2014.6736748.

[16] S. Kilaru, H. K, S. T, A. C. L and B. T, "Review and analysis of promising technologies with respect to Fifth generation networks," 2014 First International Conference on Networks & Soft Computing (ICNSC2014), Guntur, 2014, pp. 248-251, doi: 10.1109/CNSC.2014.6906653 F. Khan and Z. Pi, "mmWave mobile broadband (MMB): Unleashing the 3–300GHz spectrum," 34th IEEE Sarnoff Symposium, Princeton, NJ, 2011, pp. 1-6, doi: 10.1109/SARNOF.2011.5876482.

[17] F. Khan and Z. Pi, "mmWave mobile broadband (MMB): Unleashing the 3–300GHz spectrum," 34th IEEE Sarnoff Symposium, Princeton, NJ, 2011, pp. 1-6, doi: 10.1109/SARNOF.2011.5876482.

[18] G. Fettweis and S. Alamouti, "5G: Personal mobile internet beyond what cellular did to telephony," in IEEE Communications Magazine, vol. 52, no. 2, pp. 140-145, February 2014, doi: 10.1109/MCOM.2014.6736754.

[19] I. Parvez, A. Rahmati, I. Guvenc, A. I. Sarwat and H. Dai, "A Survey on Low Latency Towards 5G: RAN, Core Network and Caching Solutions," in IEEE Communications Surveys & Tutorials, vol. 20, no. 4, pp. 3098-3130, Fourthquarter 2018, doi: 10.1109/COMST.2018.2841349.

[20] M. Gidlund, T. Lennvall and J. Åkerberg, "Will 5G become yet another wireless technology for industrial automation?," 2017 IEEE International Conference on Industrial Technology (ICIT), Toronto, ON, 2017, pp. 1319-1324, doi: 10.1109/ICIT.2017.7915554.

[21] E. K. Aseri, "A Pragmatic Evaluation of 4G and 5G Wireless Networks in the Current Scenario," 2019 2nd International Conference on Intelligent Communication and Computational Techniques (ICCT), Jaipur, India, 2019, pp. 1-7, doi: 10.1109/ICCT46177.2019.8968780.

[22] "FANTASTIC-5G: 5G-PPP Project on 5G Air Interface Below 6 GHz." [Online]. Available: http://fantastic5g.com/wp-content/uploads/2015/07/EuCNC-FANTASTIC-5G_final.pdf

[23] A. Samuylov et al., "Characterizing Spatial Correlation of Blockage Statistics in Urban mmWave Systems," 2016 IEEE Globecom Workshops (GC Wkshps), Washington, DC, 2016, pp. 1-7, doi: 10.1109/GLOCOMW.2016.7848859.

[24] K. Pedersen, F. Frederiksen, G. Berardinelli and P. Mogensen, "A Flexible Frame Structure for 5G Wide Area," 2015 IEEE 82nd Vehicular Technology Conference (VTC2015-Fall), Boston, MA, 2015, pp. 1-5, doi: 10.1109/VTCFall.2015.7390791.

[25] J. B. Rao and A. O. Fapojuwo, "On the Tradeoff Between Spectral Efficiency and Energy Efficiency of Homogeneous Cellular Networks With Outage Constraint," in IEEE Transactions on Vehicular Technology, vol. 62, no. 4, pp. 1801-1814, May 2013, doi: 10.1109/TVT.2012.2235867.

[26] L. Shen and K. Feng, "Millimeter Wave Multiuser Beam Clustering and Iterative Power Allocation Schemes," 2019 IEEE 90th Vehicular Technology Conference (VTC2019-Fall), Honolulu, HI, USA, 2019, pp. 1-5, doi: 10.1109/VTCFall.2019.8891280.

[27] T. E. Bogale and L. B. Le, "Massive MIMO and mmWave for 5G Wireless HetNet: Potential Benefits and Challenges," in IEEE Vehicular Technology Magazine, vol. 11, no. 1, pp. 64-75, March 2016, doi: 10.1109/MVT.2015.2496240.

[28] Z. Pi and F. Khan, "A millimeter-wave massive MIMO system for next generation mobile broadband," 2012 Conference Record of the Forty Sixth Asilomar Conference on Signals, Systems and Computers (ASILOMAR), Pacific Grove, CA, 2012, pp. 693-698, doi: 10.1109/ACSSC.2012.6489100.

[29] J. Nam, J. Ahn, A. Adhikary and G. Caire, "Joint spatial division and multiplexing: Realizing massive MIMO gains with limited channel state information," 2012 46th Annual Conference on Information Sciences and Systems (CISS), Princeton, NJ, 2012, pp. 1-6, doi: 10.1109/CISS.2012.6310934.

[30] E. G. Larsson, O. Edfors, F. Tufvesson and T. L. Marzetta, "Massive MIMO for next generation wireless systems," in IEEE Communications Magazine, vol. 52, no. 2, pp. 186-195, February 2014, doi: 10.1109/MCOM.2014.6736761.

[31] W. Roh et al., "Millimeter-wave beamforming as an enabling technology for 5G cellular communications: theoretical feasibility and prototype results," in IEEE Communications Magazine, vol. 52, no. 2, pp. 106-113, February 2014, doi: 10.1109/MCOM.2014.6736750.

[32] L. Jiao, N. Wang, P. Wang, A. Alipour-Fanid, J. Tang and K. Zeng, "Physical Layer Key Generation in 5G Wireless Networks," in IEEE Wireless Communications, vol. 26, no. 5, pp. 48-54, October 2019, doi: 10.1109/MWC.001.1900061.

[33] R. Q. Hu and Y. Qian, "An energy efficient and spectrum efficient wireless heterogeneous network framework for 5G systems," in IEEE Communications Magazine, vol. 52, no. 5, pp. 94-101, May 2014, doi: 10.1109/MCOM.2014.6815898.

[34] S. Nagul, "A review on 5G modulation schemes and their comparisons for future wireless communications," 2018 Conference on Signal Processing And Communication Engineering Systems (SPACES), Vijayawada, 2018, pp. 72-76, doi: 10.1109/SPACES.2018.8316319.

[35] Y. Cai, Z. Qin, F. Cui, G. Y. Li and J. A. McCann, "Modulation and Multiple Access for 5G Networks," in IEEE Communications Surveys

& Tutorials, vol. 20, no. 1, pp. 629-646, Firstquarter 2018, doi: 10.1109/COMST.2017.2766698.

[36] M. Agiwal, A. Roy and N. Saxena, "Next Generation 5G Wireless Networks: A Comprehensive Survey," in IEEE Communications Surveys & Tutorials, vol. 18, no. 3, pp. 1617-1655, thirdquarter 2016, doi: 10.1109/COMST.2016.2532458.

[37] M. N. Tehrani, M. Uysal and H. Yanikomeroglu, "Device-to-device communication in 5G cellular networks: challenges, solutions, and future directions," in IEEE Communications Magazine, vol. 52, no. 5, pp. 86-92, May 2014, doi: 10.1109/MCOM.2014.6815897.

[38] C. Liang, F. R. Yu and X. Zhang, "Information-centric network function virtualization over 5g mobile wireless networks," in IEEE Network, vol. 29, no. 3, pp. 68-74, May-June 2015, doi: 10.1109/MNET.2015.7113228.

[39] B. A. A. Nunes, M. Mendonca, X. Nguyen, K. Obraczka and T. Turletti, "A Survey of Software-Defined Networking: Past, Present, and Future of Programmable Networks," in IEEE Communications Surveys & Tutorials, vol. 16, no. 3, pp. 1617-1634, Third Quarter 2014, doi: 10.1109/SURV.2014.012214.00180.

[40] A. A. Barakabitze, A. Ahmad, R. Mijumbi, and A. Hines, “5G network slicing using SDN and NFV: A survey of taxonomy, architectures and future challenges,” Computer Networks, vol. 167, p. 106984, 2020.

[41] “Gartner Says 8.4 Billion Connected ‘Things’ Will Be in Use in 2017, Up 31 Percent From 2016,” Gartner. [Online]. Available: https://www.gartner.com/en/newsroom/press-releases/2017-02-07-gartner-says-8-billion-connected-things-will-be-in-use-in-2017-up-31-percent-from-2016

[42] “Cellular Networks for Massive IoT - Enabling Low Power Wide Area Applications,” Internet of Things. [Online]. Available: https://www.gsma.com/iot/resources/cellular-networks-for-massive-iot-enabling-low-power-wide-area-applications/

[43] S. Li, L. D. Xu, and S. Zhao, “5G Internet of Things: A survey,” Journal of Industrial Information Integration, 20-Feb-2018. [Online]. Available: https://www.sciencedirect.com/science/article/abs/pii/S2452414X18300037

[44] A. Checko et al., "Cloud RAN for Mobile Networks—A Technology Overview," in IEEE Communications Surveys & Tutorials, vol. 17, no. 1, pp. 405-426, Firstquarter 2015, doi: 10.1109/COMST.2014.2355255.

[45] C. Tsai and M. Moh, "Load balancing in 5G cloud radio access networks supporting IoT communications for smart communities," 2017 IEEE International Symposium on Signal Processing and Information Technology (ISSPIT), Bilbao, 2017, pp. 259-264, doi: 10.1109/ISSPIT.2017.8388652.

[46] D. Wang, Y. Wang, R. Sun and X. Zhang, "Robust C-RAN Precoder Design for Wireless Fronthaul with Imperfect Channel State Information," 2017 IEEE Wireless Communications and Networking Conference (WCNC), San Francisco, CA, 2017, pp. 1-6, doi: 10.1109/WCNC.2017.7925528.

[47] K. Liolis, A. Geurtz, R. Sperber, D. Schulz, S. Watts, G. Poziopoulou, B. Evans, N. Wang, O. Vidal, B. T. Jou, M. Fitch, S. D. Sendra, P. S. Khodashenas, and N. Chuberre, “Use cases and scenarios of 5G integrated satellite-terrestrial networks for enhanced mobile broadband: The SaT5G approach,” International Journal of Satellite Communications and Networking, vol. 37, no. 2, pp. 91–112, 2018.

[48] P. Popovski, K. F. Trillingsgaard, O. Simeone and G. Durisi, "5G Wireless Network Slicing for eMBB, URLLC, and mMTC: A Communication-Theoretic View," in IEEE Access, vol. 6, pp. 55765-55779, 2018, doi: 10.1109/ACCESS.2018.2872781.

[49] M. Alsenwi, N. H. Tran, M. Bennis, A. Kumar Bairagi and C. S. Hong, "eMBB-URLLC Resource Slicing: A Risk-Sensitive Approach," in IEEE Communications Letters, vol. 23, no. 4, pp. 740-743, April 2019, doi: 10.1109/LCOMM.2019.2900044.

[50] S. M. A. Kazmi et al., "Resource management in dense heterogeneous networks," 2015 17th Asia-Pacific Network Operations and Management Symposium (APNOMS), Busan, 2015, pp. 440-443, doi: 10.1109/APNOMS.2015.7275383.

[51] P. Rost et al., "Network Slicing to Enable Scalability and Flexibility in 5G Mobile Networks," in IEEE Communications Magazine, vol. 55, no. 5, pp. 72-79, May 2017, doi: 10.1109/MCOM.2017.1600920.

[52] J. Ordonez-Lucena, P. Ameigeiras, D. Lopez, J. J. Ramos-Munoz, J. Lorca and J. Folgueira, "Network Slicing for 5G with SDN/NFV: Concepts, Architectures, and Challenges," in IEEE Communications Magazine, vol. 55, no. 5, pp. 80-87, May 2017, doi: 10.1109/MCOM.2017.1600935.

[53] Z. Kotulski et al., "On end-to-end approach for slice isolation in 5G networks. Fundamental challenges," 2017 Federated Conference on Computer Science and Information Systems (FedCSIS), Prague, 2017, pp. 783-792, doi: 10.15439/2017F228.

[54] I. Afolabi, T. Taleb, K. Samdanis, A. Ksentini and H. Flinck, "Network Slicing and Softwarization: A Survey on Principles, Enabling Technologies, and Solutions," in IEEE Communications Surveys & Tutorials, vol. 20, no. 3, pp. 2429-2453, thirdquarter 2018, doi: 10.1109/COMST.2018.2815638.

[55] D. T. Hoang, D. Niyato, P. Wang, A. De Domenico and E. C. Strinati, "Optimal Cross Slice Orchestration for 5G Mobile Services," 2018 IEEE 88th Vehicular Technology Conference (VTC-Fall), Chicago, IL, USA, 2018, pp. 1-5, doi: 10.1109/VTCFall.2018.8690608.

[56] Z. Feng, C. Qiu, Z. Feng, Z. Wei, W. Li and P. Zhang, "An effective approach to 5G: Wireless network virtualization," in IEEE Communications Magazine, vol. 53, no. 12, pp. 53-59, Dec. 2015, doi: 10.1109/MCOM.2015.7355585.

[57] C. Liang and F. R. Yu, "Wireless Network Virtualization: A Survey, Some Research Issues and Challenges," in IEEE Communications Surveys & Tutorials, vol. 17, no. 1, pp. 358-380, Firstquarter 2015, doi: 10.1109/COMST.2014.2352118.

[58] R. Irmer et al., "Coordinated multipoint: Concepts, performance, and field trial results," in IEEE Communications Magazine, vol. 49, no. 2, pp. 102-111, February 2011, doi: 10.1109/MCOM.2011.5706317.

[59] J. G. Andrews, "Seven ways that HetNets are a cellular paradigm shift," in IEEE Communications Magazine, vol. 51, no. 3, pp. 136-144, March 2013, doi: 10.1109/MCOM.2013.6476878.

[60] G. Barish and K. Obraczke, "World Wide Web caching: trends and techniques," in IEEE Communications Magazine, vol. 38, no. 5, pp. 178-184, May 2000, doi: 10.1109/35.841844.

[61] L. Li, G. Zhao and R. S. Blum, "A Survey of Caching Techniques in Cellular Networks: Research Issues and Challenges in Content Placement and Delivery Strategies," in IEEE Communications Surveys & Tutorials, vol. 20, no. 3, pp. 1710-1732, thirdquarter 2018, doi: 10.1109/COMST.2018.2820021.

[62] N. Golrezaei, A. F. Molisch, A. G. Dimakis and G. Caire, "Femtocaching and device-to-device collaboration: A new architecture for wireless video distribution," in IEEE Communications Magazine, vol. 51, no. 4, pp. 142-149, April 2013, doi: 10.1109/MCOM.2013.6495773.

[63] P. Mach and Z. Becvar, "Mobile Edge Computing: A Survey on Architecture and Computation Offloading," in IEEE Communications Surveys & Tutorials, vol. 19, no. 3, pp. 1628-1656, thirdquarter 2017, doi: 10.1109/COMST.2017.2682318.

[64] N. Abbas, Y. Zhang, A. Taherkordi and T. Skeie, "Mobile Edge Computing: A Survey," in IEEE Internet of Things Journal, vol. 5, no. 1, pp. 450-465, Feb. 2018, doi: 10.1109/JIOT.2017.2750180.

[65] Y. Mao, C. You, J. Zhang, K. Huang and K. B. Letaief, "A Survey on Mobile Edge Computing: The Communication Perspective," in IEEE Communications Surveys & Tutorials, vol. 19, no. 4, pp. 2322-2358, Fourthquarter 2017, doi: 10.1109/COMST.2017.2745201.

[66] D. Niyato and E. Hossain, "Radio resource management games in wireless networks: an approach to bandwidth allocation and admission control for polling service in IEEE 802.16 [Radio Resource Management and Protocol Engineering for IEEE 802.16]," in *IEEE Wireless Communications*, vol. 14, no. 1, pp. 27-35, Feb. 2007.

[67] X. Kang, R. Zhang and M. Motani, "Price-Based Resource Allocation for Spectrum-Sharing Femtocell Networks: A Stackelberg Game Approach," in IEEE Journal on Selected Areas in Communications, vol. 30, no. 3, pp. 538-549, April 2012, doi: 10.1109/JSAC.2012.120404.

[68] P. Semasinghe, E. Hossain and K. Zhu, "An Evolutionary Game for Distributed Resource Allocation in Self-Organizing Small Cells," in IEEE Transactions on Mobile Computing, vol. 14, no. 2, pp. 274-287, 1 Feb. 2015, doi: 10.1109/TMC.2014.2318700.

[69] L. Song, Y. Li, Z. Ding and H. V. Poor, "Resource Management in

Non-Orthogonal Multiple Access Networks for 5G and Beyond," in IEEE Network, vol. 31, no. 4, pp. 8-14, July-August 2017, doi: 10.1109/MNET.2017.1600287.

[70] X. Zhu, C. Jiang, L. Kuang, Z. Zhao and S. Guo, "Two-Layer Game Based Resource Allocation in Cloud Based Integrated Terrestrial-Satellite Networks," in IEEE Transactions on Cognitive Communications and Networking, vol. 6, no. 2, pp. 509-522, June 2020, doi: 10.1109/TCCN.2020.2981016.

[71] P. Semasinghe, S. Maghsudi and E. Hossain, "Game Theoretic Mechanisms for Resource Management in Massive Wireless IoT Systems," in IEEE Communications Magazine, vol. 55, no. 2, pp. 121-127, February 2017, doi: 10.1109/MCOM.2017.1600568CM.

[72] C. Yang, J. Li, M. Sheng, A. Anpalagan and J. Xiao, "Mean Field Game-Theoretic Framework for Interference and Energy-Aware Control in 5G Ultra-Dense Networks," in IEEE Wireless Communications, vol. 25, no. 1, pp. 114-121, February 2018, doi: 10.1109/MWC.2017.1600114.

[73] K. Hamidouche, W. Saad, M. Debbah and H. V. Poor, "Mean-Field Games for Distributed Caching in Ultra-Dense Small Cell Networks," 2016 American Control Conference (ACC), Boston, MA, 2016, pp. 4699-4704, doi: 10.1109/ACC.2016.7526096.

[74] H. Kim, J. Park, M. Bennis, S. Kim and M. Debbah, "Mean-Field Game Theoretic Edge Caching in Ultra-Dense Networks," in IEEE Transactions on Vehicular Technology, vol. 69, no. 1, pp. 935-947, Jan. 2020, doi: 10.1109/TVT.2019.2953132.

[75] S. Ranadheera, S. Maghsudi and E. Hossain, "Minority Games With Applications to Distributed Decision Making and Control in Wireless Networks," in IEEE Wireless Communications, vol. 24, no. 5, pp. 184-192, October 2017, doi: 10.1109/MWC.2017.1600351WC.

[76] M. Elmachkour, E. Sabir, A. Kobbane, J. Ben-Othman and M. El koutbi, "The greening of spectrum sensing: a minority game-based mechanism design," in IEEE Communications Magazine, vol. 52, no. 12, pp. 150-156, December 2014, doi: 10.1109/MCOM.2014.6979967..

[77] P. Mähönen and M. Petrova, "Minority Game for Cognitive Radios: Cooperating without Cooperation," Physical Commun., vol. 1, no. 2, 2008, pp. 94–102.

[78] F. Y. Lin, C. Hsiao, Y. Wen and Y. Wu, "Optimization-Based Resource Management Strategies for 5G C-RAN Slicing Capabilities," 2018 Tenth International Conference on Ubiquitous and Future Networks (ICUFN), Prague, 2018, pp. 346-351, doi: 10.1109/ICUFN.2018.8436837.

[79] M. Jiang, M. Condoluci and T. Mahmoodi, "Network slicing management & prioritization in 5G mobile systems," European Wireless 2016; 22th European Wireless Conference, Oulu, Finland, 2016, pp. 1-6.

[80] N. Zhang, Y. Liu, H. Farmanbar, T. Chang, M. Hong and Z. Luo, "Network Slicing for Service-Oriented Networks Under Resource Constraints," in IEEE Journal on Selected Areas in Communications, vol. 35, no. 11, pp. 2512-2521, Nov. 2017, doi: 10.1109/JSAC.2017.2760147.

[81] A. Kammoun, N. Tabbane, G. Diaz, A. Dandoush and N. Achir, "End-to-End Efficient Heuristic Algorithm for 5G Network Slicing," 2018 IEEE 32nd International Conference on Advanced Information Networking and Applications (AINA), Krakow, 2018, pp. 386-392, doi: 10.1109/AINA.2018.00065.

[82] K. Shanmugam, N. Golrezaei, A. G. Dimakis, A. F. Molisch and G. Caire, "FemtoCaching: Wireless Content Delivery Through Distributed Caching Helpers," in IEEE Transactions on Information Theory, vol. 59, no. 12, pp. 8402-8413, Dec. 2013, doi: 10.1109/TIT.2013.2281606.

[83] M. Dehghan et al., "On the Complexity of Optimal Request Routing and Content Caching in Heterogeneous Cache Networks," in IEEE/ACM Transactions on Networking, vol. 25, no. 3, pp. 1635-1648, June 2017, doi: 10.1109/TNET.2016.2636843.

[84] Y.-C. Chang, S.-Y. Huang, and H.-C. Chao, "Metaheuristic-Based Scheme for Spectrum Resource Schedule Over 5G IoT Network," Lecture Notes of the Institute for Computer Sciences, Social Informatics and Telecommunications Engineering IoT as a Service, pp. 117–125, 2018.

[85] A. L. Rezaabad, H. Beyranvand, J. A. Salehi and M. Maier, "Ultra-Dense 5G Small Cell Deployment for Fiber and Wireless Backhaul-Aware Infrastructures," in IEEE Transactions on Vehicular Technology, vol. 67, no. 12, pp. 12231-12243, Dec. 2018, doi: 10.1109/TVT.2018.2875114.

[86] P. Zhou, J. Xu, W. Wang, C. Jiang, K. Wang and J. Hu, "Human-Behavior and QoE-Aware Dynamic Channel Allocation for 5G Networks: A Latent Contextual Bandit Learning Approach," in IEEE Transactions on Cognitive Communications and Networking, vol. 6, no. 2, pp. 436-451, June 2020, doi: 10.1109/TCCN.2020.2969631.

[87] C. Zhang, M. Dong and K. Ota, "Fine-Grained Management in 5G: DQL Based Intelligent Resource Allocation for Network Function Virtualization in C-RAN," in IEEE Transactions on Cognitive Communications and Networking, vol. 6, no. 2, pp. 428-435, June 2020, doi: 10.1109/TCCN.2020.2982886.

[88] Y. Li, W. Zhang, C. Wang, J. Sun and Y. Liu, "Deep Reinforcement Learning for Dynamic Spectrum Sensing and Aggregation in Multi-Channel Wireless Networks," in IEEE Transactions on Cognitive Communications and Networking, vol. 6, no. 2, pp. 464-475, June 2020, doi: 10.1109/TCCN.2020.2982895.

[89] L. Yu et al., "Spectrum Availability Prediction for Cognitive Radio Communications: A DCG Approach," in IEEE Transactions on Cognitive Communications and Networking, vol. 6, no. 2, pp. 476-485, June 2020, doi: 10.1109/TCCN.2020.2973572.

[90] Y. Lu, L. Liu, J. Panneerselvam, B. Yuan, J. Gu and N. Antonopoulos, "A GRU-Based Prediction Framework for Intelligent Resource Management at Cloud Data Centres in the Age of 5G," in IEEE Transactions on Cognitive Communications and Networking, vol. 6, no. 2, pp. 486-498, June 2020, doi: 10.1109/TCCN.2019.2954388.

[91] D. Zhao, H. Qin, B. Song, Y. Zhang, X. Du and M. Guizani, "A Reinforcement Learning Method for Joint Mode Selection and Power Adaptation in the V2V Communication Network in 5G," in IEEE Transactions on Cognitive Communications and Networking, vol. 6, no. 2, pp. 452-463, June 2020, doi: 10.1109/TCCN.2020.2983170.

[92] M. Chen, Y. Miao, H. Gharavi, L. Hu and I. Humar, "Intelligent Traffic Adaptive Resource Allocation for Edge Computing-Based 5G Networks," in IEEE Transactions on Cognitive Communications and Networking, vol. 6, no. 2, pp. 499-508, June 2020, doi: 10.1109/TCCN.2019.2953061.

[93] M. Bunyakitanon, X. Vasilakos, R. Nejabati and D. Simeonidou, "End-to-End Performance-Based Autonomous VNF Placement With Adopted Reinforcement Learning," in IEEE Transactions on Cognitive Communications and Networking, vol. 6, no. 2, pp. 534-547, June 2020, doi: 10.1109/TCCN.2020.2988486.

[94] J. Hall, K. Moessner, R. MacKenzie, F. Carrez and C. H. Foh, "Dynamic Scheduler Management Using Deep Learning," in IEEE Transactions on Cognitive Communications and Networking, vol. 6, no. 2, pp. 575-585, June 2020, doi: 10.1109/TCCN.2020.2980529.

[95] J. Luo, F. R. Yu, Q. Chen and L. Tang, "Adaptive Video Streaming With Edge Caching and Video Transcoding Over Software-Defined Mobile Networks: A Deep Reinforcement Learning Approach," in IEEE Transactions on Wireless Communications, vol. 19, no. 3, pp. 1577-1592, March 2020, doi: 10.1109/TWC.2019.2955129.

[96] F. Guo, F. R. Yu, H. Zhang, H. Ji, M. Liu and V. C. M. Leung, "Adaptive Resource Allocation in Future Wireless Networks With Blockchain and Mobile Edge Computing," in IEEE Transactions on Wireless Communications, vol. 19, no. 3, pp. 1689-1703, March 2020, doi: 10.1109/TWC.2019.2956519.

[97] J. Feng, F. Richard Yu, Q. Pei, X. Chu, J. Du and L. Zhu, "Cooperative Computation Offloading and Resource Allocation for Blockchain-Enabled Mobile-Edge Computing: A Deep Reinforcement Learning Approach," in IEEE Internet of Things Journal, vol. 7, no. 7, pp. 6214-6228, July 2020, doi: 10.1109/JIOT.2019.2961707.

[98] J. Feng, F. R. Yu, Q. Pei, J. Du and L. Zhu, "Joint Optimization of Radio and Computational Resources Allocation in Blockchain-Enabled Mobile Edge Computing Systems," in IEEE Transactions on Wireless Communications, vol. 19, no. 6, pp. 4321-4334, June 2020, doi: 10.1109/TWC.2020.2982627.

[99] Y. Liu, F. R. Yu, X. Li, H. Ji and V. C. M. Leung, "Decentralized Resource Allocation for Video Transcoding and Delivery in Blockchain-Based System With Mobile Edge Computing," in IEEE Transactions on Vehicular Technology, vol. 68, no. 11, pp. 11169-

11185, Nov. 2019, doi: 10.1109/TVT.2019.2937351.

[100] D. Wu, "QoS provisioning in wireless networks," Wireless Communications and Mobile Computing, vol. 5, no. 8, pp. 957–969, 2005.

[101] N. U. Ginige, K. B. Shashika Manosha, N. Rajatheva and M. Latva-aho, "Admission Control in 5G Networks for the Coexistence of eMBB-URLLC Users," 2020 IEEE 91st Vehicular Technology Conference (VTC2020-Spring), Antwerp, Belgium, 2020, pp. 1-6, doi: 10.1109/VTC2020-Spring48590.2020.9129141.

[102] T. Sigwele, P. Pillai, A. S. Alam, and Y. F. Hu, "Fuzzy logic-based call admission control in 5G cloud radio access networks with preemption," *EURASIP Journal on Wireless Communications and Networking*, vol. 2017, no. 1, 2017.

[103] R. K. Gupta and R. Misra, "Machine Learning-based Slice allocation Algorithms in 5G Networks," 2019 International Conference on Advances in Computing, Communication and Control (ICAC3), Mumbai, India, 2019, pp. 1-4, doi: 10.1109/ICAC347590.2019.9036741.

[104] M. R. Raza, C. Natalino, P. Öhlen, L. Wosinska and P. Monti, "A Slice Admission Policy Based on Reinforcement Learning for a 5G Flexible RAN," 2018 European Conference on Optical Communication (ECOC), Rome, 2018, pp. 1-3, doi: 10.1109/ECOC.2018.8535483.

[105] T. V. K. Buyakar, H. Agarwal, B. R. Tamma and A. A. Franklin, "Resource Allocation with Admission Control for GBR and Delay QoS in 5G Network Slices," 2020 International Conference on COMmunication Systems & NETworkS (COMSNETS), Bengaluru, India, 2020, pp. 213-220, doi: 10.1109/COMSNETS48256.2020.9027310.

[106] J. Feng, Q. Pei, F. R. Yu, X. Chu, J. Du and L. Zhu, "Dynamic Network Slicing and Resource Allocation in Mobile Edge Computing Systems," in IEEE Transactions on Vehicular Technology, vol. 69, no. 7, pp. 7863-7878, July 2020, doi: 10.1109/TVT.2020.2992607.

[107] R. Trivisonno, Q. Wei and C. C. Marquezan, "QoS Enhancements for V2X Services in 5G Networks," 2020 IEEE 91st Vehicular Technology Conference (VTC2020-Spring), Antwerp, Belgium, 2020, pp. 1-5, doi: 10.1109/VTC2020-Spring48590.2020.9129065.

[108] C. V. Murudkar and R. D. Gitlin, "Optimal-Capacity, Shortest Path Routing in Self-Organizing 5G Networks using Machine Learning," 2019 IEEE 20th Wireless and Microwave Technology Conference (WAMICON), Cocoa Beach, FL, USA, 2019, pp. 1-5, doi: 10.1109/WAMICON.2019.8765434.

[109] U. Chunduri, A. Clemm and R. Li, "Preferred Path Routing - A Next-Generation Routing Framework beyond Segment Routing," 2018 IEEE Global Communications Conference (GLOBECOM), Abu Dhabi, United Arab Emirates, 2018, pp. 1-7, doi: 10.1109/GLOCOM.2018.8647410.

[110] A. A. Barakabitze, L. Sun, I. Mkwawa and E. Ifeachor, "A Novel QoE-Centric SDN-Based Multipath Routing Approach for Multimedia Services over 5G Networks," 2018 IEEE International Conference on Communications (ICC), Kansas City, MO, 2018, pp. 1-7, doi: 10.1109/ICC.2018.8422617.

[121]

[111] T. Chi, C. Lin, J. Kuo and W. Chen, "Live Video Multicast for Dynamic Users via Segment Routing in 5G Networks," 2018 IEEE Global Communications Conference (GLOBECOM), Abu Dhabi, United Arab Emirates, 2018, pp. 1-7, doi: 10.1109/GLOCOM.2018.8647865.

[112] M. Bagaa, D. L. C. Dutra, T. Taleb and K. Samdanis, "On SDN-Driven Network Optimization and QoS Aware Routing Using Multiple Paths," in IEEE Transactions on Wireless Communications, vol. 19, no. 7, pp. 4700-4714, July 2020, doi: 10.1109/TWC.2020.2986408.

[113] R. Li, K. Makhijani and L. Dong, "New IP: A Data Packet Framework to Evolve the Internet : Invited Paper," 2020 IEEE 21st International Conference on High Performance Switching and Routing (HPSR), Newark, NJ, USA, 2020, pp. 1-8, doi: 10.1109/HPSR48589.2020.9098996.

[114] J. Li and X. Zhang, "Deep Reinforcement Learning Based Joint Scheduling of eMBB and URLLC in 5G Networks," in IEEE Wireless Communications Letters, doi: 10.1109/LWC.2020.2997036.

[115] D. Panno and S. Riolo, "A New Joint Scheduling Scheme for GBR and non-GBR Services in 5G RAN," 2018 IEEE 23rd International Workshop on Computer Aided Modeling and Design of Communication Links and Networks (CAMAD), Barcelona, 2018, pp. 1-6, doi: 10.1109/CAMAD.2018.8514964.

[116] M. Yan, G. Feng, J. Zhou, Y. Sun and Y. Liang, "Intelligent Resource Scheduling for 5G Radio Access Network Slicing," in IEEE Transactions on Vehicular Technology, vol. 68, no. 8, pp. 7691-7703, Aug. 2019, doi: 10.1109/TVT.2019.2922668.

[117] D. Luong, A. Outtagarts and Y. Ghamri-Doudane, "Multi-level Resource Scheduling for network slicing toward 5G," 2019 10th International Conference on Networks of the Future (NoF), Rome, Italy, 2019, pp. 25-31, doi: 10.1109/NoF47743.2019.9015028.

[118] W. U. Rehman, T. Salam, A. Almogren, K. Haseeb, I. Ud Din and S. H. Bouk, "Improved Resource Allocation in 5G MTC Networks," in IEEE Access, vol. 8, pp. 49187-49197, 2020

[119] H. S. Ben Abdelmula, M. N. M. Warip, O. B. Lynn and N. Yaakob, "An efficient scheduling scheme for heterogeneous services in OFDMA based 5G LTE-Advanced network with carrier aggregation," 2018 IEEE Symposium on Computer Applications & Industrial Electronics (ISCAIE), Penang, 2018, pp. 275-280, doi: 10.1109/ISCAIE.2018.8405484.

[120] Y. Liu, F. R. Yu, X. Li, H. Ji, and V. C. M. Leung, "Distributed self-optimizing interference management in ultra-dense networks with non-orthogonal multiple access," *Wireless Networks*, vol. 26, no. 4, pp. 2809–2823, 2020.